\documentclass[aps, amsfonts, nofootinbib,superscriptaddress]{revtex4}
\usepackage{amsmath}
\usepackage{graphicx}
\usepackage{float}
\usepackage{dsfont}
\usepackage{xcolor}

\newcommand{\simge}{\hspace*{0.2em}\raisebox{0.5ex}{$>$}
     \hspace{-0.8em}\raisebox{-0.3em}{$\sim$}\hspace*{0.2em}}

\newcommand{\figref}[1]{Fig.~\ref{#1}}

\newcommand{\bea}{\begin{eqnarray}}
\newcommand{\eea}{\end{eqnarray}}

\newcommand{\Lag}{{\mathcal L}}

\begin{document}

\title{Nonrelativistic Effective Field Theory with a Resonance Field}

\author{J.~B.~Habashi}
\affiliation{Department of Physics, University of Arizona, Tucson, AZ 85721, USA} 
\author{S.~Fleming}
\affiliation{Department of Physics, University of Arizona, Tucson, AZ 85721, USA}  
\author{U.~van~Kolck} 
\affiliation{Universit\'e Paris-Saclay, CNRS/IN2P3, IJCLab, 91405 Orsay, France}
\affiliation{Department of Physics, University of Arizona, Tucson, AZ 85721, USA} 

\begin{abstract}
We discuss shallow resonances in the nonrelativistic scattering of
two particles using an effective field theory (EFT) 
that includes an auxiliary field
with the quantum numbers of the resonance.
We construct the manifestly renormalized scattering amplitude up to 
next-to-leading order in a systematic expansion.
For a narrow resonance, the amplitude is perturbative 
except in the immediate vicinity of the resonance poles.
It naturally has a zero in the low-energy region,
analogous to the Ramsauer-Townsend effect.
For a broad resonance, the leading-order 
amplitude is nonperturbative almost everywhere
in the regime of validity of the EFT.
We regain the results of an EFT without the 
auxiliary field, which is equivalent to the effective-range
expansion with large scattering length and effective range.
We also consider an additional fine tuning leading to a low-energy 
amplitude zero
even for a broad resonance.
We show that in all cases the requirement of renormalizability
when the auxiliary field is not a ghost
ensures the resonance poles are in the lower half of the complex
momentum plane, as expected by other arguments.
The systematic character of the EFT expansion is exemplified with
a toy model serving as underlying theory.
\end{abstract}

\date{\today}

\maketitle

\section{Introduction}
\label{intro}

Information about quantum mechanical systems comes from 
two-body scattering experiments,
a prominent feature of which are resonance peaks. 
Resonances typically reflect interactions that are not strong enough
to produce a bound or virtual state, as for 
nucleon-alpha particle and alpha-alpha scattering. 
When the resonance is visible at energies
that are small compared to those of the break-up of the scattering
particles, we can describe the reaction through an effective field theory
(EFT) with those particles as degrees of freedom
--- for a recent, comprehensive review, see Ref. \cite{Hammer:2019poc}.

A resonance peak can be associated with a pole in the $S$ matrix
near the positive energy axis of the complex energy plane. 
An EFT aims for a model-independent description of the $S$ matrix
at low energies, which is constrained only by symmetries.
A shallow state is associated with a momentum much smaller than the 
inverse of the range of the interaction and requires only contact interactions. 
Including all possible 
contact interactions with an arbitrary number of derivatives
ensures that the EFT can describe the low-energy $S$ matrix of a finite-range
interaction without the need to know the exact form of
the interaction at short distances.
Maintaining model independence at the quantum-mechanical level
demands renormalization, that is, insensitivity
to the regularization needed to yield finite observables.
By their very nature, $S$-matrix poles are non-perturbative
and require a summation of Feynman diagrams to all loop orders 
for a subset of interactions.
The challenge for EFTs that include shallow poles is to 
produce a renormalized and realistic leading-order amplitude
while treating subleading interactions in a 
systematic, perturbative expansion.

The EFT for a single shallow pole on the positive or negative imaginary axis
in the complex momentum plane --- representing, respectively,
a bound or virtual state --- is well understood 
\cite{vanKolck:1997ut,Kaplan:1998tg,Kaplan:1998we,vanKolck:1998bw}.
This EFT applies when the two-body scattering length is much larger
than the effective range, such
as for nucleons or $^4$He atoms --- for an introduction, 
see Ref. \cite{vanKolck:2019vge}.
These are the simplest ``halo'' states --- intrinsically quantum-mechanical
states with size larger than the range of the underlying interaction.
The EFT for shallow resonances, which can be thought of as unbound halo states,
has not been as fully developed.

The first formulation of an EFT for nonrelativistic resonances was proposed
\cite{Bertulani:2002sz,Bedaque:2003wa}
for neutron-alpha $p$-wave scattering.
It was argued that at least two parameters are needed for renormalization,
which was carried out with an auxiliary ``dimer'' field \cite{Kaplan:1996nv}
having the quantum numbers of the resonance, 
the ground state of the $^5$He nucleus.
The theory was generalized to narrow resonances 
--- those very close to real energies --- in any partial wave 
in Ref. \cite{Bedaque:2003wa}
and applied, in the presence of the Coulomb interaction,
to the $s$-wave resonance (the ground state
of the $^8$Be nucleus) in alpha-alpha scattering in Ref. \cite{Higa:2008dn}.
For the case of the Delta resonance in 
Compton scattering on the nucleon, a similar idea was implemented
independently in Ref. \cite{Pascalutsa:2002pi}
and reformulated along the lines of 
Ref. \cite{Bedaque:2003wa} for $p$-wave pion-nucleon scattering 
in Ref. \cite{Long:2009wq}.
The importance of the non-resonant background in 
the description of narrow $s$-wave resonances was emphasized
in Ref. \cite{Gelman:2009be},
with subleading corrections further investigated in 
Ref. \cite{Alhakami:2017ntb}.

The dimer field is quite useful
as the energy-dependent interaction
it produces is equivalent to the resummation of a subset 
of contact interactions. Recently we presented an equivalent
momentum-dependent description of an $s$-wave resonance 
without the dimer field \cite{Habashi:2020qgw}.
It accounts well for two low-energy poles produced by a generic
short-range potential, reproducing the first two terms 
in the effective-range expansion with scattering length and effective range
of comparable size.
This includes a broad resonance
represented by a pair of poles of the $S$ matrix that do not lie close to 
the real axis and are sometimes not considered ``true'' resonances.
It does not naturally accommodate a narrow resonance and 
its background.

Here we extend the EFT for $s$-wave resonances to cover 
both broad and narrow resonances with a dimer field.
Our formalism applies to
a variety of situations where shallow $s$-wave
resonances appear, ranging from atomic to nuclear
to particle physics. 
In atomic physics, the resonance behavior of 
neutral atoms is
often studied with models closely related to the EFT presented here
\cite{Braaten:2007nq}.
In nuclear physics, 
shallow $s$-wave resonances can be seen in low-energy neutron scattering 
\cite{Vogt:1962zz,Gunsing:2018pvl}.
With the inclusion of the Coulomb interaction, many resonances in 
proton-nucleus or nucleus-nucleus scattering
would also fall within the scope of the EFT,
as long as the resonance's characteristic size is 
larger than the nucleus'.
In particle physics, some of the many exotic hadronic states containing
heavy quarks \cite{Guo:2017jvc}
are $s$-wave resonances.
An example is the $\Lambda_c^+(2595)$ just above the 
threshold for break up into $\Sigma_c^+(2455)$ and $\pi^0$ \cite{Zyla:2020zbs},
although nearby charged thresholds need to be accounted for as well.

After a brief description of the EFT with a dimer field in
Sec. \ref{dimerEFT}, we show in Sec. \ref{broadresonance} 
that the dimer formulation of broad $s$-wave resonances 
gives indeed the same on-shell results 
as momentum-dependent interactions \cite{Habashi:2020qgw}.
In Sec. \ref{narrowresonance} we revisit the case of narrow resonances
and point out that the background introduced in Ref. \cite{Gelman:2009be} 
leads to sufficiently strong energy dependence for the 
scattering amplitude to admit a zero in the low-energy region.
We also correct the renormalization procedure of a
calculation found in the literature \cite{Alhakami:2017ntb}.
Section \ref{toy} is dedicated to a toy model that illustrates
some of the aspects of the EFT for narrow resonances.
In Sec. \ref{ampzero} the case is considered of an additional fine tuning
leading to an amplitude zero in presence of a broad resonance. 
We conclude in Sec. \ref{conclusion}.

\section{EFT with a resonance field}
\label{dimerEFT}

The central paradigm of EFT is separation of scales: low-energy degrees of 
freedom (at a scale $M_{lo}$) cannot resolve high-energy physics
(at a scale $M_{hi}$).
The latter is encoded in 
the parameters --- ``low-energy constants'' (LECs) --- of 
all the interactions among the low-energy degrees of 
freedom which are allowed by symmetries.
The degrees of freedom in EFT are fields that incorporate the
creation and annihilation of particles.
The simplest way to account for symmetries is through a Lagrangian,
whose infinite number of terms are organized according to the magnitude
of their effects on observables.
This ``power counting'' (PC) underlines the expansion 
of observables in powers of the ratio $Q/M_{hi}$, where 
$Q \sim M_{lo}\ll M_{hi}$ is the characteristic external momentum of a process.
We refer to successive terms in the expansion in $Q/M_{hi}$ as 
leading order (LO), next-to-leading order (NLO), {\it etc.}
The EFT breaks down at $M_{hi}$.

We construct the most general two-body Lagrangian which 
for simplicity we assume to be invariant under 
time reversal, parity, and Lorentz transformations,
limiting ourselves to a single stable particle species.
At energies below the particle mass $m$, Lorentz invariance is most
easily implemented in a $Q/m$ expansion, which gives rise to a 
nonrelativistic expansion. 
Pair production cannot occur and particle number is conserved. 
One can use a field $\psi$ that involves only the annihilation of particles.
The Lagrangian is Hermitian and all LECs are real. 
At very low energies the dominant partial wave is $s$, 
and we restrict ourselves to this wave here. Generalization to other
waves is straightforward but tedious. 
We are interested in the case where there is a shallow resonance
and introduce a dimer auxiliary field $d$ \cite{Weinberg:1962hj,Kaplan:1996nv}
with its quantum numbers
and a residual mass $\Delta$.
The most general Lagrangian is then
\bea
\Lag & = & \psi^\dagger \left(i\frac{\partial}{\partial t} 
+ \frac{\vec{\nabla}^2}{2 m} \right) \psi 
+ d^\dagger \left( i\frac{\partial}{\partial t} 
+ \frac{\vec{\nabla}^2}{4 m} - \Delta \right) d 
\notag \\ 
& & 
+ \sqrt{\frac{4 \pi}{m}} \, g_{0} 
\left( d^\dagger \psi \psi + \psi^\dagger\psi^\dagger d \right)  
- \frac{4 \pi}{m} C_{0} \left(\psi^\dagger \psi\right)\left(\psi^\dagger \psi\right)
\notag \\ 
& & 
+ \sqrt{\frac{4 \pi}{m}}\, \frac{g_{2}}{4} 
\left[ d^\dagger \left(\psi \overleftrightarrow{\nabla}^2 \psi \right) 
+ \left( \psi \overleftrightarrow{\nabla}^2 \psi \right)^{\dagger} d \right] 
+ \frac{4\pi}{m} \frac{C_{2}}{8} 
\left[\left( \psi \psi \right)^{\dagger} 
\left(\psi \overleftrightarrow{\nabla}^2 \psi \right) 
+ \left(\psi \overleftrightarrow{\nabla}^2 \psi \right)^{\dagger} 
\left(\psi \psi \right) \right] 
\notag \\ 
& & 
+ \ldots \,,
\label{eq.new1}
\eea
where $C_{2n}$ and $g_{2n}$ are real LECs
and ``$\ldots$'' indicates terms with additional fields or derivatives. 
There is a certain redundancy in the Lagrangian \eqref{eq.new1},
since when one integrates out $d$ one obtains a string of
four-$\psi$ interactions of the form already present in $\Lag$
\cite{Bedaque:1999vb}.
However, these terms are all correlated in a way that is explicit
only once the dimer field is kept. The four-field interactions retained
in Eq. \eqref{eq.new1} can be thought of as the uncorrelated part of these 
interactions.
In other situations one might need to capture a different correlation
by making the dimer a ghost field with a negative kinetic term 
\cite{Kaplan:1996nv}.

With the standard rules of quantum field theory the Lagrangian \eqref{eq.new1}
leads to an infinite number of contributions to the $T$ matrix.
We want to organize these contributions at energy $E\equiv k^2/m$
in an expansion
\begin{equation}
T(k) = T^{(0)}(k) + T^{(1)} (k)
+ \ldots 
\label{eq.new2} 
\end{equation}
or, alternatively,
\begin{equation}
\frac{1}{T(k)} = \frac{1}{T^{(0)}(k)}
\left(1- \frac{T^{(1)}(k)}{T^{(0)}(k)} 
+ \ldots \right)\,, 
\label{eq.new3}
\end{equation}
where $T^{(n)}$ represents a term of relative ${\cal O}(Q^n/M_{hi}^n)$.
The $S$ matrix is then obtained from
\begin{equation}
S(k)=1-\frac{imk}{2\pi}\, T(k) 
\, .
\label{eq.new5}
\end{equation}
Note that, except at $k=0$, the poles of $S$ and $T$ are the same. 
In the cases of interest here, $S(k)$ has $N=2,3$ (complex) poles 
denoted by $k_{n}$, and it can be written as 
({\it cf.} Refs. \cite{Hu:1948zz,Peierls:1959,Newton:1960})
\begin{equation}
S(k) = 
(-1)^N \, \exp{(2i\phi(k))} \,
\prod\limits_{n=1}^{N} \frac{k - k_{n}^{\ast}}{k - k_{n}} \, ,
\label{eq.new86}
\end{equation}
where $\phi(k)$ is a smooth background phase.
We are particularly concerned with shallow resonances, 
which consist of a pair of poles 
in the complex momentum plane,
\begin{equation}
k_{\pm} = \pm k_{R} - i k_{I} \, , 
\label{eq.new4}
\end{equation}
with real $k_{R,I}>0$.
In the absence of other shallow poles,
the $S$ matrix can be expressed in terms
of the resonance energy and (energy-dependent) width,
respectively
\begin{equation}
E_R = \frac{1}{m} \left(k_R^2+k_I^2\right) \, ,
\qquad
\Gamma(E) = 4k_I \sqrt{\frac{E}{m}} \, ,
\label{eq.new6}
\end{equation}
as
\begin{equation}
S(E) = \exp\left(2i\phi(E)\right) \,
\frac{E-E_R-i\Gamma(E)/2}{E-E_R+i\Gamma(E)/2} 
\,.
\label{eq.new8}
\end{equation}

To obtain $T$, one first needs to regulate the theory.
We choose the conceptually simplest regulator, a cutoff $\Lambda$
in momentum space.
Results should be independent of regulator choice, which is achieved
by the process of renormalization. At each order in the expansion,
positive powers of $\Lambda$ arising from the loops 
should be removed from observables
by the cutoff dependence of a finite number of bare LECs in $\Lag$. 
This fixes the cutoff dependence of the LEC at that order.
The finite combination of bare LECs and loops that remains
is then fitted to an equal number of inputs.
When the underlying theory is known, one can match its results
for a certain number of observables. 
When the underlying theory is not known or cannot be calculated,
one can use experimental data instead.
In either case, since at each order only a finite number of
LECs is present, observables not used in the fit can be predicted. 
Higher-order contributions will eventually affect not only these observables
but also observables that were already used as input at lower orders.
When that happens a LEC that was fixed needs to be changed.
It is therefore convenient to split a LEC as
\begin{equation}
\alpha(\Lambda) = \alpha^{(0)}(\Lambda) + \alpha^{(1)}(\Lambda) + \ldots \,,
\label{eq.new9}
\end{equation}
where $\alpha^{(n)}(\Lambda)$ is fixed at order $n$ in the expansion.

This procedure is standard and applied to any EFT --- for example,
applications to nuclear physics can be found in Ref. \cite{Hammer:2019poc}.
What depends on the physical situation is the PC that determines
which interactions appear at each order.
From naive dimensional analysis (NDA) 
--- for a review see Ref. \cite{vanKolck:2020plz} ---
we expect that $\Delta= \mathcal{O}(M_{hi}^2/m)$,
$g_{2n} = \mathcal{O}(m^{-1/2}M_{hi}^{1/2-n})$,
and $C_{2n} = \mathcal{O}(M_{hi}^{-2n-1})$. 
In this case the expansion \eqref{eq.new2} is obtained from strict
perturbation theory and there are no poles within the regime
of applicability of the EFT. 

Below we consider the various deviations from NDA that can give rise
to a low-energy resonance, $E_R\ll {\cal O}(M_{hi}^2/m)$.
This can be achieved with
\begin{equation}
\Delta^{(0)} = \mathcal{O}\left(\frac{M_{lo}^2}{m}\right) 
\label{eq.new10}
\end{equation}
because then both kinetic and residual mass terms
of the dimer are expected to have similar sizes for momenta 
$Q={\cal O}(M_{lo})$.
Already the bare dimer propagator,
\begin{equation}
D^{(0)}_0(k) = \frac{1}{k^2/m - \Delta^{(0)} + i\epsilon} \,,
\label{eq.new11}
\end{equation}
displays two poles. With two particle legs attached at each end, 
$4\pi g_0^{(0)2}D^{(0)}_0(k)/m$ is an energy-dependent potential.
Whether this potential can be treated in perturbation theory 
or instead quantum corrections are important and move the poles significantly 
will depend on the scaling of $g_0$.
The two parameters $\Delta^{(0)}$ and $g_0^{(0)}$ will determine
the energy $E_{R}$ and decay width $\Gamma$ of the resonance.

The first quantum correction $D_1(k)$
to the bare dimer propagator \eqref{eq.new11}
consists of the one-loop dimer self-energy, that is, an insertion of a 
particle bubble,
\begin{equation}
\frac{D_1(k)}{D_0^{(0)}(k)} =  -g_0^{(0)2} D_0^{(0)}(k) \, I_0(k)\,,
\label{eq.new12}
\end{equation}
where the loop gives rise to
\bea
I_{0} & = & L_{1} + i k + L_{-1}k^2 + L_{-3}k^4 + \ldots  \,,
\label{eq.new13} 
\\
L_{n} & \equiv &
\theta_{n} \, \Lambda^{n} \,.
\label{eq.new14}
\eea
Here $\theta_{n}$ are pure numbers 
whose values depend on the regularization scheme. 
For a sharp momentum cutoff, for example, $\theta_{n}= 2 /n \pi$.

The cutoff-dependent terms are eventually removed by renormalization.
As in any EFT, the loop contribution that is not affected by a LEC
is the non-analytic term, here the $ik$ in Eq. \eqref{eq.new13}.
Thus it gives an estimate of the relative importance of $D_1(k)$,
\begin{equation}
\frac{D_1(k)}{D_0^{(0)}(k)} = {\cal O}\left(\frac{m g_0^{(0)2}}{M_{lo}}\right)\,,
\label{eq.new15}
\end{equation}
for momenta $Q\sim M_{lo}$.
Additional loop insertions in between bare dimer propagators
bring in additional powers of this factor.
The size of $g_0^{(0)}$ determines whether the series of loops in 
the dimer propagator needs to be resummed or not,
leading to a broad or narrow resonance, respectively.
In all of the PCs we consider, we assume an additional suppression
in interactions with LECs $g_{2n>0}$ relative to $g_0$, which is given by NDA,
\begin{equation}
g_{2n>0}={\cal O}\left(\frac{g_0}{M_{hi}^{2n}}\right) \,,
\label{eq.new16}
\end{equation}
so they enter at N$^2$LO or higher.
In calculations up to NLO, we see no renormalization enhancement 
over this estimate.
If such an enhancement is found at higher orders, it can easily be accounted
for.

As we will see, another interesting feature of the $T$ matrix can appear
depending on the scaling of $C_0$: a point on the real momentum axis where
the amplitude vanishes, which is analogous to the Ramsauer-Townsend
effect \cite{Ramsauer:1921,Townsend:1922}.
The scaling of the LECs $C_{2n>0}$ will depart from NDA if
the scaling of $C_0$ does. When $C_0$ satisfies NDA, we assume $C_{2n>0}$
does as well.

\section{Broad resonance}
\label{broadresonance}

In certain regions of parameter space, 
a potential consisting of an attractive well surrounded by
a repulsive barrier of range $R$ displays a pair of low-energy poles
with complex momenta of magnitude $|k|\ll R^{-1}$.
If the attraction is not too strong, the two poles are at 
momenta \eqref{eq.new4} with
$k_I\sim k_R ={\cal O}(M_{lo})>0$.
In this case the resonance width and energy are comparable
$\Gamma \sim E_R ={\cal O}(M_{lo}^2/m)$.

\begin{figure}[t]
\includegraphics[trim={3.15cm 24.25cm 3.00cm 2.15cm},clip]{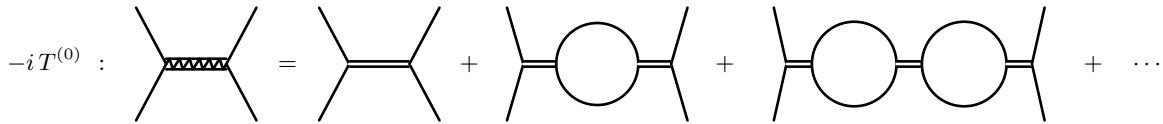}
\caption{Leading-order $T$ matrix for a broad resonance in terms
of the dressed dimer propagator (shaded double line) 
arising from the resummation of diagrams with successive 
particle (solid line) bubble insertions in 
a bare dimer propagator (double line).}
\label{fig.1}
\end{figure}

Since the imaginary part of the complex pole momentum comes from loop 
integration, regardless of the detailed form of the underlying interactions
we must have
\begin{equation}
g_{0}^{(0)} = \mathcal{O}\left(\sqrt{\frac{M_{lo}}{m}}\right) \,. 
\label{eq.new17}
\end{equation}
At LO we then need to resum the one-loop dimer self-energy 
as in \figref{fig.1}, resulting in the dressed dimer propagator
\begin{equation}
D^{(0)}(k) = \frac{D^{(0)}_0(k)}{1 + g_{0}^{(0)2} D^{(0)}_0(k) \, I_{0}(k)}\,.
\label{eq.new18}
\end{equation}
This propagator contains the small scale $M_{lo}$ associated with
a low-energy resonance. 

Here we assume that $C_0$ is not enhanced with respect to NDA,
\begin{equation}
C_{0}^{(0)} = 0 \,,
\qquad 
C_{0}^{(1)} = \mathcal{O}\left(\frac{1}{M_{hi}}\right)
\, .
\label{eq.new19}
\end{equation}
This scaling arises naturally in a simple model \cite{Habashi:2020qgw}.
With this assumption the LO $T$ matrix is simply
\bea
T^{(0)}(k) & = & \frac{4 \pi}{m}\,g_{0}^{(0)2}D^{(0)}(k) 
=-\frac{4 \pi}{m}\left(-\frac{1}{a_{0}} + \frac{r_0}{2}\, k^2
- ik - L_{-1} k^2 + \ldots\right)^{-1}
\,.
\label{eq.new20}
\eea
In Eq.~\eqref{eq.new20} we have absorbed the cutoff dependence from 
$L_1$ (which diverges if $\Lambda\to \infty$) 
in the bare LEC $\Delta^{(0)}(\Lambda)$, while $g_{0}^{(0)}$ does not run
with the cutoff. 
We have defined the renormalized parameters $a_0^{-1}$
and $r_0$ through
\bea
\Delta^{(0)}(\Lambda) & = & -\frac{2}{m r_{0}}\left(L_{1}-\frac{1}{a_{0}}\right)
\,,
\label{eq.new21} \\
g_{0}^{(0)} & = & \left(-\frac{2}{m r_{0}}\right)^{1/2} \,. 
\label{eq.new22}
\eea
The remaining cutoff dependence in Eq.~\eqref{eq.new20} can be made
arbitrarily small by taking the cutoff arbitrarily large.
For a cutoff $\Lambda\simge M_{hi}$, those terms are no larger
than higher-order terms. We are free to neglect this residual cutoff
dependence at LO, as it will be removed by LECs appearing at 
higher orders.
What we have done here is to carry out the procedure 
of Ref. \cite{Bertulani:2002sz} in the $s$ wave.
Equation~\eqref{eq.new20} has the form of the effective-range expansion 
truncated at the level of the effective range.
The parameters $a_0$ and $r_0$ are nothing but the standard
scattering length and effective range, respectively.
They are obtained by fitting the underlying or empirical
amplitude at two momenta.
With the assumed scalings \eqref{eq.new10} and \eqref{eq.new17}, 
they are both large in the sense that they are set by the low-energy scale,
$|a_0|\sim |r_0| ={\cal O}(1/M_{lo})$.
While Eq. \eqref{eq.new21} allows for any sign of $a_0$,
Eq. \eqref{eq.new22} requires $r_0< 0$.

The same LO amplitude was obtained in a formulation without the dimer field
where the no- and two-derivative contact interactions have a
different scaling than here, which demands their resummation 
\cite{Habashi:2020qgw}.
Renormalization could only be implemented for $r_0< 0$.
The resulting pole structure was discussed in detail.
In particular, $r_0< 0$ constrains resonance poles to be in the lower
half of the complex momentum plane with
\bea
k_{I}^{(0)} & = & -\frac{1}{r_{0}} >0 \, ,
\label{eq.new23} \\
k_{R}^{(0)} & = & - \frac{1}{r_{0}}\sqrt{\frac{2r_{0}}{a_0} - 1}>0
\, ,
\label{eq.new24}
\eea
for $2r_{0}<a_{0}<0$.
In the limit $\Lambda\to \infty$, Eq. \eqref{eq.new20} can be rewritten as
Eq. \eqref{eq.new8} with
\begin{equation}
\phi^{(0)}(E) = 0 \,.
\label{eq.new25}
\end{equation}
From the assumed PC,
$k_I^{(0)}\sim k_R^{(0)} ={\cal O}(M_{lo})$,
or alternatively $\Gamma^{(0)}\sim E_R^{(0)} ={\cal O}(M_{lo}^2/m)$.

Resonance pole positions in the lower half-plane 
are in agreement with the general requirement on the $S$ matrix
\cite{Moller:1946,Hu:1948zz,Schuetzer:1951} 
that leads to states decaying with time. 
The constraint $r_0<0$ in Eq. \eqref{eq.new20} allows also for
two purely imaginary poles, one of which
is a virtual state on the negative imaginary axis and the other,
either a virtual state (for $a_{0}<2r_{0}$) 
or a bound state on the positive imaginary axis (for $a_{0}>0$) 
\cite{Habashi:2020qgw}. 
Other possibilities, thought to be unphysical, are excluded
by renormalization.

Here, the pole location constraint arises from the standard, positive
sign of the kinetic term of the dimer in the 
Lagrangian \eqref{eq.new1}. 
At the two-body level there seems to be no {\it a priori}
restriction on the sign of the dimer kinetic term 
\cite{Kaplan:1996nv,Beane:2000fi}.
When this kinetic term is treated perturbatively, as
is the case when there is a single low-energy pole (bound or virtual state),
either sign seems to be allowed in many-body calculations \cite{Hammer:2019poc}.
The dimer in this situation involves no correlation
among the four-field interactions of a no-dimer formulation.
When the kinetic term is treated nonperturbatively
for a positive effective range, however,
the additional two-body pole 
has negative residue.
The associated negative probability 
leads to problems beyond elastic two-body scattering,
for example in the three-body system where 
the pole lies in the region of integration \cite{Gabbiani:2000hr}.

The remainder of the inverse of the LO $T$ matrix,
\begin{equation}
\delta T^{(1)-1}(k) \equiv \frac{m}{4\pi} \left(L_{-1} k^2 + \ldots\right)
\,,
\label{eq.new26}
\end{equation}
indicates that corrections enter at NLO, with $g_{0}^{(1)}$
adjusted to keep the effective range unchanged. 
However, $g_{0}^{(1)}$ induces cutoff dependence in the scattering length,
which can be corrected with $\Delta^{(1)}$. This leads to the scaling
\begin{equation}
g_{0}^{(1)} = \mathcal{O}\left(\frac{M_{lo}}{M_{hi}}\sqrt{\frac{M_{lo}}{m}}\right)
\,, 
\qquad
\Delta^{(1)} = \mathcal{O}\left(\frac{M_{lo}^3}{mM_{hi}}\right) 
\,.
\label{eq.new27}
\end{equation}
NDA implies $C_{0}^{(1)}$, Eq. \eqref{eq.new19}, enters at this order as well.

\begin{figure}[t]
\includegraphics[trim={4.1cm 24.2cm 4.1cm 2.15cm},clip]{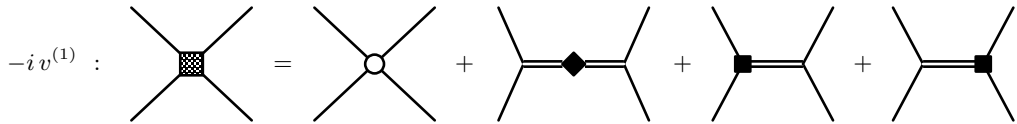}
\caption{Next-to-leading-order potential from
the four-particle contact interaction (empty circle),
dimer residual mass (solid diamond), 
and two-particle/dimer coupling (solid square).
Other symbols as in \figref{fig.1}.}
\label{fig.2}
\end{figure}

\begin{figure}[t]
\includegraphics[trim={4.0cm 24.2cm 4.2cm 2.15cm},clip]{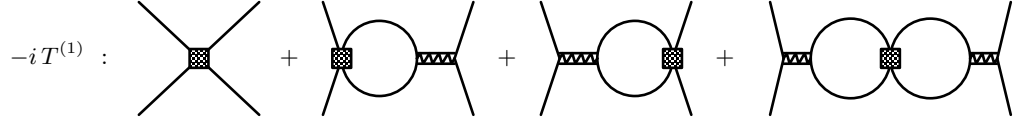}
\caption{Next-to-leading-order $T$ matrix for a broad resonance
in terms of the NLO potential (shaded square, \figref{fig.2})
and the LO $T$ matrix (shaded double line, \figref{fig.1}).}
\label{fig.3}
\end{figure}

\begin{figure}[t]
\includegraphics[trim={3.3cm 22.45cm 3.45cm 2.15cm},clip]{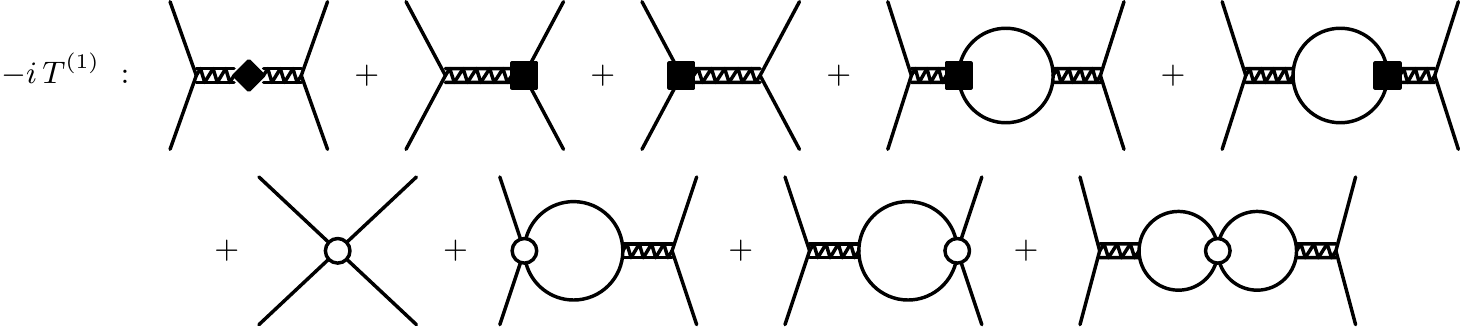}
\caption{Extended version of the next-to-leading-order $T$ matrix for 
a broad resonance in \figref{fig.3}. 
Notation as in Figs. \ref{fig.1} and \ref{fig.2}.}
\label{fig.4}
\end{figure}

These contributions add to a correction
in the energy-dependent potential, displayed in \figref{fig.2}.
The NLO potential is then inserted once in  
diagrams with LO interactions as sketched in \figref{fig.3}.
A more explicit drawing of the corresponding diagrams can be found
in \figref{fig.4}.
This procedure, also known as
first-order distorted-wave perturbation theory,
leads to
\bea
\frac{4\pi}{m} \biggl(\frac{T^{(1)}}{T^{(0)2}} - \delta T^{(1)-1}\biggr) &=&
C_{0}^{(1)} \left(L_{1}-\frac{1}{a_0}\right)^2 
-g_{0}^{(1)} \left(L_{1}-\frac{1}{a_0}\right) \sqrt{-2mr_{0}} 
-\Delta^{(1)} \, \frac{mr_{0}}{2} 
\notag \\
&& +\left[C_{0}^{(1)}\left(L_{1}-\frac{1}{a_0}\right)  
- g_{0}^{(1)} \sqrt{- \frac{mr_{0}}{2}}
-\frac{L_{-1}}{r_0}\right] r_{0} k^2  
+ C_{0}^{(1)} \,\frac{r_{0}^2k^4}{4} - L_{-3} k^4 + \ldots
\notag \\
&=& -P_0 \left(\frac{r_0}{2}\right)^3 k^4 + \ldots
\,.
\label{eq.new28}
\eea
The four-particle contact interaction $C_{0}^{(1)}$ introduces a
$k^4$ dependence, which leads to a non-vanishing shape parameter 
$|P_0|={\cal O}(M_{lo}/M_{hi})$.
This additional parameter at NLO requires fitting the underlying or empirical
amplitude at an additional momentum.
At the same time, 
the linear cutoff-dependence in $L_{1}$ appears in the other two terms
in Eq. \eqref{eq.new28},
which means that in addition to $C_{0}^{(1)}$ we indeed need both $\Delta^{(1)}$ 
and $g_{0}^{(1)}$ for renormalization. In Eq. \eqref{eq.new28} we imposed
that the LO renormalized parameters $a_0$ and $r_0$ were not changed,
leading to the running of $\Delta^{(1)}$ and $g_{0}^{(1)}$: 
\bea
C_{0}^{(1)} & = & - \frac{r_0 P_0}{2} \,,
\label{eq.new29}\\
g_{0}^{(1)}(\Lambda) & = & -\left(-\frac{2}{m r_{0}}\right)^{1/2}
\left[\frac{r_0 P_0}{2} \left(L_1-\frac{1}{a_0}\right)+\frac{L_{-1}}{r_{0}}\right]
\,,
\label{eq.new30}\\
\Delta^{(1)}(\Lambda) & = & \frac{2}{mr_{0}} \left(L_{1}-\frac{1}{a_0}\right)  
\left[\frac{r_0 P_0}{2}\left(L_1-\frac{1}{a_0}\right)+\frac{2L_{-1}}{r_{0}}\right] 
\,.
\label{eq.new31}
\eea
Again $g_{0}^{(1)}(\Lambda)$ is real due to the constraint $r_0<0$ 
stemming from a positive dimer kinetic term. 

The amplitude up to NLO can
be written as 
\begin{equation}
T^{(0+1)}(k)= -\frac{4 \pi}{m}\left[-\frac{1}{a_{0}} + \frac{r_{0}}{2}\,k^2
-P_0 \left(\frac{r_0}{2}\right)^3 k^4 - ik \right]^{-1} + \ldots \,.
\label{eq.new32}
\end{equation}
Again, this amplitude was obtained from momentum-dependent interactions
in Ref. \cite{Habashi:2020qgw}.
The NLO correction \eqref{eq.new28} is perturbative within the regime of 
applicability of the EFT, so it does not lead to new poles.
Our choice of keeping LO renormalized parameters unchanged 
does mean the poles are slightly displaced from their LO positions,
as discussed in Ref. \cite{Habashi:2020qgw}:
\bea
k_{I}^{(0+1)} & = & -\frac{1}{r_{0}} 
\left[1+ P_0\left(\frac{r_0}{a_0}-1\right)\right]>0 \, ,
\label{eq.new33} \\
k_{R}^{(0+1)} & = & - \frac{1}{r_{0}}\sqrt{\frac{2r_{0}}{a_0} - 1}
\left[1+ P_0\left(1-\frac{r_0^2/2a_0^2}{2r_0/a_0-1}\right)\right]>0 
\, .
\label{eq.new34}
\eea
The NLO correction amounts to
a background \cite{Hu:1948zz,vanKampen:1953}
proportional to $P_0$: again up to higher-order terms,
the $S$ matrix can be written as Eq. \eqref{eq.new8} with 
\cite{Habashi:2020qgw}
\begin{equation}
\phi^{(0+1)}(E) = \frac{P_0r_0}{2} \sqrt{mE}\,.
\label{eq.new35}
\end{equation}

The ``\ldots'' in Eq. \eqref{eq.new32}
contain terms $\propto \Lambda^{-3}$ indicating that
new interactions appear at or before N$^3$LO.
For example, from NDA we expect $g_2$ and $C_2$ to appear at 
N$^2$LO and N$^3$LO, respectively.
These corrections can be incorporated following the procedure
described above, although expressions become lengthier as
we have to consider multiple insertions of subleading 
interactions --- for example, two insertions of NLO interactions
at N$^2$LO.
We refrain from this exercise here, shifting instead to a
qualitatively different case.

\section{Narrow resonance}
\label{narrowresonance}

A narrow resonance is one for which $\Gamma\ll E_{R}$,
or equivalently $k_I\ll k_R$.
In this case the quantum corrections to the bare dimer propagator
are generically small. 
In zeroth order in perturbation theory,
the poles lie on the real axis, with 
the $ik$ from Eq. \eqref{eq.new12} a correction to Eq. \eqref{eq.new11}
that requires no resummation, except in a small window of energies
around the resonance \cite{Bedaque:2003wa}.
In Sec. \ref{narrowresonancegenmom} we consider generically low 
momenta, while the region around the resonance 
is tackled in Sec. \ref{narrowresonancewindow}.

\subsection{Generic momenta}
\label{narrowresonancegenmom}

We can account for a narrow resonance by adjusting the PC of the previous
section. Since the contribution from loop integrals does not change,
we need a new scaling for $g_{0}^{(0)}$ \cite{Bedaque:2003wa},
\begin{equation}
g_{0}^{(0)} = \mathcal{O}\left(\sqrt{\frac{M_{lo}^2}{m M_{hi}}}\right) \,. 
\label{eq.new36}
\end{equation}
Each loop comes with a factor of $Q/M_{hi}$ and 
is therefore perturbative for generic momenta $Q\sim M_{lo}$. 
A side effect of the smallness of $g_{0}^{(0)}$
is that a natural $C_0$,
\begin{equation}
C_{0}^{(0)}=\mathcal{O}\left(\frac{1}{M_{hi}}\right) \,,
\label{eq.new37}
\end{equation}
now appears at LO \cite{Gelman:2009be}.

\begin{figure}[t]
\includegraphics[trim={6.7cm 24.25cm 6.7cm 2.15cm},clip]{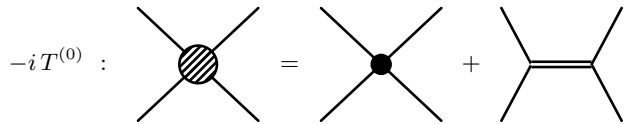}
\caption{Leading-order $T$ matrix 
in the presence of a narrow resonance for generic momenta,
in terms
of the leading four-particle contact interaction (filled circle) 
and the bare dimer propagator (double line).}
\label{fig.5}
\end{figure}

The LO $T$ matrix now is simply the sum of tree-level diagrams
--- the bare dimer propagator
\eqref{eq.new11} and the four-particle contact interaction ---
shown in \figref{fig.5}. That is,
\begin{equation}
T^{(0)}(k) = \frac{4\pi}{m}\left(C_{0}^{(0)} + g_{0}^{(0)2} D^{(0)}_0(k) \right)  
= \frac{4\pi}{m}\, a_0 \, \frac{k^2/k_{0}^2-1}{k^2/k_r^{2}-1} \,,
\label{eq.new38} 
\end{equation}
where we defined
\bea 
\Delta^{(0)} & = & \frac{k_r^{2}}{m} \,,
\label{eq.new39}
\\
g_{0}^{(0)} & = & \left[a_0 \left(\frac{k_r^{2}}{k_{0}^2}-1\right)
\frac{k_r^{2}}{m}\right]^{1/2}  \,,
\label{eq.new40}
\\
C_{0}^{(0)} & = & \frac{k_r^{2}}{k_{0}^2} \, a_0\,. 
\label{eq.new41}
\eea
The amplitude \eqref{eq.new38} has real poles,
\bea
k_I^{(0)} &=& 0 \, , 
\label{eq.new42}
\\
k_R^{(0)} &=& k_r >0\, , 
\label{eq.new43}
\eea
with $k_r= {\cal O}(M_{lo})$, for $\Delta^{(0)} > 0$. 
($\Delta^{(0)} < 0$, on the other hand, generates a bound/virtual state pair.
We do not consider this case explicitly, although there is no obvious
obstacle to do so.)
As we discuss in detail in Sec. \ref{narrowresonancewindow}, 
corrections to Eq. \eqref{eq.new38} are large in the immediate
neighborhood of these poles.
In addition, the presence of $C_{0}^{(0)}$ at LO 
together with the dimer field leads to amplitude zeros at $k=\pm k_0$,
$|k_0|= {\cal O}(M_{lo})$.
For $\Delta^{(0)} > g_0^{(0)2}/C_0^{(0)}$, 
a zero $k_0> 0$ occurs on the scattering axis.
In the particular case $k_0= 0$ --- that is, 
$g_{0}^{(0)2} = C_{0}^{(0)} \Delta^{(0)}$ ---
the $T$ matrix vanishes at threshold,
which means that the scattering length $a_0=0$.
More generally, $|a_0|={\cal O}(1/M_{hi})$ but the sign of $a_0$
is constrained by the relative
sizes of $k_r^2$ and $k_0^2$, 
\begin{equation}
a_0  k_r^{2} \left(\frac{k_r^{2}}{k_{0}^2}-1\right) > 0 \,,
\label{eq.new44}
\end{equation}
so that $g_{0}^{(0)}$ in Eq. \eqref{eq.new40} is real.

The dimer part of the $T$ matrix \eqref{eq.new38} was put forward
in Ref. \cite{Bedaque:2003wa}. Reference \cite{Gelman:2009be}
was the first to point out that $C_0$ should be included as well 
on the basis of NDA, but its role in producing an amplitude zero
was not mentioned.
In contrast to the case of a broad resonance, in general 
three non-vanishing parameters need to be determined at LO, 
which requires data at three different momenta.
One can be taken as the location $k_{0}$ of the $T$-matrix zero.
The positions of the poles will be displaced at NLO
and only their real part $k_R$ can be used as input at LO.
As a third input we can take, for example, the amplitude at $k=0$,
that is, the scattering length $a_0$. 

For the present case there is no residual cutoff dependence at LO
and no need for new interactions at NLO. However,
the one-loop diagrams will bring in cutoff dependence,
requiring for renormalization NLO shifts in the three LECs already
present at LO:
\begin{equation}
g_{0}^{(1)} = \mathcal{O}\left(\frac{M_{lo}}{M_{hi}}\sqrt{\frac{M_{lo}^2}{mM_{hi}}}
\right) \, , 
\qquad 
\Delta^{(1)} = \mathcal{O}\left(\frac{M_{lo}^3}{mM_{hi}}\right) \, , 
\qquad
C_{0}^{(1)} = \mathcal{O}\left(\frac{M_{lo}}{M_{hi}^2}\right) \,. 
\label{eq.new45}
\end{equation}

\begin{figure}[t]
\includegraphics[trim={7.6cm 24.25cm 7.7cm 2.15cm},clip]{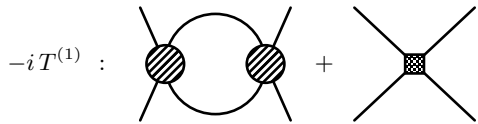}
\caption{Next-to-leading-order $T$ matrix for generic momenta
in the presence of a narrow resonance,
in terms of the one-loop correction to 
LO $T$ matrix (hatched blob, \figref{fig.5})
and the NLO potential (shaded square, \figref{fig.2}).
}
\label{fig.6}
\end{figure}

Like for a broad resonance, these corrections add to the energy-dependent
potential of \figref{fig.2}, which however must be included here
in first order of simple, non-distorted perturbation theory
along with the one-loop diagrams involving LO LECs.
The diagrams in \figref{fig.6} amount to
\begin{equation}
\frac{4\pi}{m} \frac{T^{(1)}}{T^{(0)2}} =  
\frac{C^{(1)}_{0} + 2g_{0}^{(0)}g_{0}^{(1)}D_0^{(0)} + g_{0}^{(0)2}\Delta^{(1)}D_0^{(0)2}}
{(C_{0}^{(0)} + g_{0}^{(0)2}D^{(0)}_0)^{2}}
-L_1 -ik - L_{-1}k^2 +\ldots  \,.
\label{eq.new46}
\end{equation}
The cutoff dependence from $L_1$ can be absorbed in the 
NLO bare parameters,
\bea
\Delta^{(1)}(\Lambda) & = & a_0 \left(\frac{k_r^2}{k_{0}^2}-1\right) 
\frac{k_r^2}{m} \, L_{1}
\,,
\label{eq.new47}\\
g_{0}^{(1)}(\Lambda) & = & 
-a_0\,\frac{k_r^2}{k_{0}^2}
\left[a_0\left(\frac{k_r^2}{k_{0}^2} - 1\right)\frac{k_r^2}{m}\right]^{1/2} 
\,L_{1} \,,
\label{eq.new48}\\
C_{0}^{(1)}(\Lambda) & = & \left(a_0 \, \frac{k_r^2}{k_{0}^2}\right)^2 L_{1} \,,
\label{eq.new49}
\eea
in order to keep the physical parameters that appear at LO unchanged.
Up to higher-order terms, the amplitude including NLO corrections is then
\begin{equation}
T^{(0+1)}(k)= \frac{4 \pi}{m}
\left(\frac{1}{a_0} \, \frac{k^2/k_r^{2}-1}{k^2/k_{0}^2-1} + ik + L_{-1} k^2 
+ \ldots \right)^{-1} \,,
\label{eq.new50}
\end{equation}
with residual cutoff dependence $\propto \Lambda^{-1}$.
After renormalization the only effect of NLO is to introduce
the parameter-free unitarity term $ik$.
Since $|a_0^{-1}|={\cal O}(M_{hi})$, this term is indeed relatively
small by ${\cal O}(M_{lo}/M_{hi})$ for $Q={\cal O}(M_{lo})$.
The number of LECs at LO and NLO are the same and there is no need of more 
data input at NLO. 
Expanding the denominator of Eq. \eqref{eq.new50} in powers of $k/k_0$ 
we can relate it to the effective-range expansion.
For example, the effective range is 
\begin{equation}
r_0=- \frac{2}{a_0k_r^2} \left(\frac{k_r^2}{k_0^2}-1\right)<0
\,,
\label{eq.new51}
\end{equation}
considering the constraint \eqref{eq.new44}.
Just as for a broad resonance, 
we recover the well-known constraint on the effective range
from the non-ghost character of the dimer.
In contrast to a broad resonance, however, here
$|r_0|={\cal O}(M_{hi}/M_{lo}^2)$ is large,
with $|a_0/r_0|={\cal O}(M_{lo}^2/M_{hi}^2)$. 
This is a feature of an amplitude with a low-energy zero 
\cite{vanKolck:1998bw}.
Examples where an $s$-wave zero is important
are low-energy nucleon-nucleon \cite{SanchezSanchez:2017tws}
and nucleon-deuteron \cite{Rupak:2018gnc}
scattering in, respectively, $^1S_0$ and $^2S_{1/2}$ channels,
but both appear together with virtual states instead of resonances.

The amplitude \eqref{eq.new50}
is, however, not equivalent to the effective-range expansion
truncated at the effective-range term, as is the case for a broad resonance
(see Eq. \eqref{eq.new20} and Ref. \cite{Habashi:2020qgw}).
In addition to the effective range, we obtain a string of higher powers of 
$k^2$ in the denominator of Eq. \eqref{eq.new50}
which are suppressed with respect to the effective-range term by
only ${\cal O}(Q^2/M_{lo}^2)$.
These powers, of course, resum into an amplitude zero.
In order to obtain Eq. \eqref{eq.new20} from Eq. \eqref{eq.new50}
we need an additional fine tuning to enhance $g_0$. 

A visible effect of the unitarity term is to displace the real poles 
of the LO amplitude \eqref{eq.new38} in the 
imaginary direction. The fine tuning embodied in our PC results in 
a small imaginary part $k_I$ compared to the real part,
$k_I/k_R= {\cal O}(M_{lo}/M_{hi})$:
\bea
k_I^{(0+1)} &=& \frac{a_0 k_{r}^2}{2}\left(\frac{k_{r}^2}{k_{0}^2} -1\right) >0 
\, ,
\label{eq.new52}
\\
k_R^{(0+1)} &=& k_r >0 \, . 
\label{eq.new53}
\eea
Also for a narrow resonance the location of the resonance pole
in the lower half-plane goes hand-in-hand with a negative
effective range, Eq. \eqref{eq.new51}.

The residual cutoff dependence in Eq. \eqref{eq.new50} indicates
that further corrections appear at N$^2$LO term.
Again, proceeding to higher orders is straightforward but tedious.
They will not change the qualitatively important features of LO and NLO.

\subsection{Small window around the resonance}
\label{narrowresonancewindow}

From the NLO amplitude \eqref{eq.new50} for generically low momenta
we see the emergence of a narrow pole with 
$k_I= {\cal O}(M_{lo}^2/M_{hi})\ll k_R= {\cal O}(M_{lo})$ 
or, alternatively,
$\Gamma= {\cal O}(M_{lo}^3/mM_{hi})\ll E_R= {\cal O}(M_{lo}^2/m)$.
Unfortunately the expansion of the $T$ matrix obtained
in the preceding subsection does not converge sufficiently close
to the pole \cite{Bedaque:2003wa}.
As $k$ approaches $k_R$, the LO amplitude \eqref{eq.new38} 
grows way beyond the ${\cal O}(4\pi/m M_{hi})$ magnitude 
it has for a generic $k={\cal O}(M_{lo})$,
and in fact diverges at $k=k_r$.
The first quantum correction to the bare dimer propagator diverges
twice as fast, and more loops insertions even faster.
The series stops converging in a window
$|k-k_R|\sim k_I$ around the resonance,
or in terms of energy,
$|E - E_R|\sim \Gamma$.
(Note that on account of Eq. \eqref{eq.new52}
the window size is limited by the presence of the zero, 
and if the zero is fine-tuned close to the resonance the window can be
made even narrower.)

With $g_{0}^{(0)}$ scaling as in Eq. \eqref{eq.new36} we have,
within this window around the resonance,
\begin{equation}
\left.\frac{D_1(k)}{D_0^{(0)}(k)}\right|_{\rm window} = 
{\cal O}\left(\frac{m g_{0}^{(0)2} M_{hi}}{M_{lo}^2}\right)
={\cal O}\left(1\right)\,,
\label{eq.new54}
\end{equation}
instead of Eq. \eqref{eq.new15}.
In contrast to a broad resonance, here a resummation of
the dimer propagator is necessary as a consequence of a kinematical fine tuning.
The dressed propagator is now larger than $C_0^{(0)}$ by a factor of 
$\mathcal{O}(M_{hi}/M_{lo})$.
Despite the weakness of 
$g_{0}^{(0)}$ compared to the broad resonance case, the LO $T$ matrix
is given by the full dimer propagator in \figref{fig.1} 
and requires renormalization of the residual mass
for which the $\Delta^{(0)}(\Lambda)$ in Eq. \eqref{eq.new39} 
is insufficient.
The LO $T$ matrix in the window
has the same form as Eq.~\eqref{eq.new20} 
\cite{Bedaque:2003wa,Gelman:2009be,Alhakami:2017ntb}
and can be written as
\begin{equation}
\left. T^{(-1)}(k)\right|_{\rm window} = \frac{4 \pi}{m}\,g_{0}^{(0)2}D^{(0)}(k) 
=\frac{4\pi}{m}
\left[\frac{1}{2k_{I}}\left(k^2-k_{R}^2\right)
+ik +L_{-1} k^2 + \ldots\right]^{-1}
\,,
\label{eq.new55}
\end{equation}
where, keeping the notation of the previous subsection,
\bea
\Delta^{(0)}(\Lambda) + \Delta^{(1)}(\Lambda)
& = & \frac{1}{m}\left(k_R^2+2k_IL_{1}\right)
\,,
\label{eq.new56} \\
g_{0}^{(0)} & = & \left(\frac{2k_I}{m}\right)^{1/2} \,. 
\label{eq.new57}
\eea
Apart from higher-order terms,
the $S$ matrix is again given by Eq. \eqref{eq.new8}
with no background,
\begin{equation}
\phi^{(-1)}(E) = 0 \,.
\label{eq.new58}
\end{equation}
Like for a broad resonance, it contains two parameters  
but the relations between pole parameters $k_{R,I}$ and effective
range parameters in Eqs. \eqref{eq.new23} and \eqref{eq.new24} 
no longer hold.

The NLO amplitude now receives contributions from 
$C_0^{(0)}$ (instead of $C_0^{(1)}$) as well as 
corrections to the two-particle dimer vertex and the dimer residual mass
needed for renormalization.
The NLO potential is given by \figref{fig.2}
and the NLO $T$ matrix, by \figref{fig.3} or, more explicitly, \figref{fig.4}.
These corrections 
were first considered in Ref. \cite{Gelman:2009be}, where only 
the first diagram in the second line of \figref{fig.4}
was included. 
Reference \cite{Alhakami:2017ntb} pointed out the need for
additional loop diagrams, and included the second and
third diagrams in the second line of \figref{fig.4}.
However, the last diagram in \figref{fig.4} was 
omitted even though it is of the same order.
Including all diagrams,
\bea
\left. T^{(-1+0)}(k)\right|_{\rm window} 
&=& \frac{4\pi}{m}
\left[\frac{2k_I}{k^2-k_R^2+2ik_Ik}
\left(1+\frac{2k_I^2}{k^2-k_R^2+2ik_Ik}\right)
+c\left(1-\frac{4ik_Ik}{k^2-k_R^2+2ik_Ik}\right)
\right] 
\label{eq.new59}
\\
&=& \frac{4\pi}{m}
\left\{\frac{1}{2k_{I}}\left(k^2-k_{R}^2-k_{I}^2\right)
+ik -c \left[\frac{(k^2-k_{R}^2)^2}{4k_I^2}+ k^2\right]
+ \ldots\right\}^{-1}
\,,
\label{eq.new60}
\eea
if we set
\bea
C_{0}^{(0)} & = & c \,,
\label{eq.new61}\\
g_{0}^{(1)}(\Lambda) + g_{0}^{(2)}(\Lambda) & = & -\left(\frac{2k_I}{m}\right)^{1/2}
\left[c\left(L_1 + k_I\right) + k_IL_{-1}\right]\,,
\label{eq.new62}\\
\Delta^{(2)}(\Lambda) 
& = & \frac{k_I}{m} \biggl\{2c \left[\left(L_1+k_I\right)^2+ k_R^{2}-k_I^2 \right]
+k_I+2L_{-1}\left(2k_IL_{1}+k_R^2\right)\biggr\}\,,
\label{eq.new63}
\eea
with $c$ a constant.
The cutoff-independent terms in the expressions above were chosen
so that the pole positions do not change and there is no
double pole at NLO, except for the $k_I^2$ correction in 
Eq. \eqref{eq.new59}, which accounts for the 
$k_I^2$ in Eq. \eqref{eq.new6}.
As a consequence,
the $S$ matrix just acquires a background phase 
\begin{equation}
\phi^{(-1+0)}(E) = -c\, \sqrt{mE} \,.
\label{eq.new64}
\end{equation}

Equation \eqref{eq.new59} agrees
with the corresponding result in Ref. \cite{Alhakami:2017ntb}
despite the missing diagram in the latter. 
This diagram brings in both a momentum-independent quadratic divergence
$L_1^2$ and a linear divergence $L_1$ proportional to $ik$,
in addition to a finite term proportional to $k^2$ that contributes
to the renormalization of $g_{0}^{(2)}$.
The quadratic divergence 
is absorbed in $\Delta^{(2)}$, Eq. \eqref{eq.new63},
together with a (momentum-independent) linear divergence $L_1$
from other diagrams and the residual cutoff dependence $L_{-1}$ from LO
diagrams.
The linear divergence proportional to $ik$ from the missing
diagram, being non-analytic in energy,
cannot be absorbed anywhere, but it cancels an opposite linear divergence 
from other diagrams. Other diagrams also induce additional cutoff
dependence of types $L_1$ and $L_{-1}$, which are absorbed in 
$g_{0}^{(1)}+ g_{0}^{(2)}$, Eq. \eqref{eq.new62}.
Two combinations of LECs are necessary and sufficient to remove the cutoff 
dependence from an arbitrary regulator. 
The cancellation in the non-analytic linearly divergent terms 
is absent in the incomplete set of diagrams
considered in Ref. \cite{Alhakami:2017ntb}. 
But the error was inconsequential because Ref. \cite{Alhakami:2017ntb}
used dimensional regularization with minimal subtraction,
which makes $L_1=0=L_{-1}$.
In this particular case not only does the missed diagram cause no problem,
but also no shifts in $g_0$ and $\Delta$ are needed explicitly.
In general, however, only the inclusion of
all diagrams of a given order --- here, all diagrams in \figref{fig.4} ---
leads to a renormalized result. 
Reference \cite{Alhakami:2017ntb} also looked into N$^2$LO corrections,
but again many diagrams are missing,
such as the analog of the last diagram in \figref{fig.4} with
an additional loop and $C_0$ vertex in the middle.
The additional diagrams can be included straightforwardly and 
renormalization performed following the procedure we presented above.

So far in this subsection we have written the amplitude in terms of
parameters like $k_R$, $k_I$, and $c$ that are in principle
determined from data within the resonance window. 
For simplicity, we kept the same notation for some parameters
as in the previous subsection, with the implication
that Eqs. \eqref{eq.new53}, \eqref{eq.new52}, and \eqref{eq.new41} hold.
To see that that is indeed the case, 
we rewrite Eq. \eqref{eq.new60} with only a higher-order error as
\begin{equation}
\left(\frac{m}{4\pi} T^{(-1+0)}(k)\right)_{\rm window}^{-1} =
\underbrace{\frac{1}{c}\,\frac{k^2-k_{R}^2}{k^2-(k_{R}^2-2k_I/c)}
+ik}_{{\cal O}\left(M_{lo}\right)} 
\underbrace{-\frac{k_I}{2}-ck^2}_{{\cal O}\left(\frac{M_{lo}^2}{M_{hi}}\right)}
\underbrace{+ \ldots}_{{\cal O}\left(\frac{M_{lo}^3}{M_{hi}^2}\right)} 
\,.
\label{eq.new65}
\end{equation}
To relate the parameters around the resonance 
to those outside the resonance window
we match this expression to Eq. \eqref{eq.new50},
\begin{equation}
\left(\frac{m}{4\pi} T^{(0+1)}(k)\right)^{-1} = 
\underbrace{\frac{k_{0}^2}{a_0k_r^{2}} \, 
\frac{k^2-k_r^{2}}{k^2-k_{0}^2}}_{{\cal O}\left(M_{hi}\right)}  
\underbrace{+ ik }_{{\cal O}\left(M_{lo}\right)} 
\underbrace{+ \ldots}_{{\cal O}\left(\frac{M_{lo}^2}{M_{hi}}\right)} \,.
\label{eq.new66}
\end{equation}
At the first common order, ${\cal O}(M_{lo})$,
\bea
k_R &=& k_r >0 \,, 
\label{eq.new67}
\\
k_I &=& \frac{a_0 k_{r}^2}{2}\left(\frac{k_{r}^2}{k_{0}^2} -1\right) >0 \,,
\label{eq.new68}
\\
c &=& a_0\frac{k_{r}^2}{k_{0}^2} \,,
\label{eq.new69}
\eea
in agreement with Eqs. \eqref{eq.new53}, \eqref{eq.new52}, and
\eqref{eq.new41}.
These relations plus their analogs at higher orders
ensure the consistency of the $T$ matrix
in the two regions, inside and outside of the resonance window.

\section{Toy Model}
\label{toy}

One of the important features of an EFT is its model independence, 
which means the EFT describes the low-energy limit 
of different underlying theories
as long as the separation of scales in all of them follows the same pattern. 
The details of the underlying dynamics are encoded in the renormalized values 
of the LECs, the relative importance of which is captured by PC. 
Here we exemplify the systematic character of the resulting expansions
for observables,
taking a particular potential model as an underlying theory.

The toy model we use comprises an attractive spherical well 
of range $R$ and depth $\beta^2/mR^2$ with a repulsive delta shell with 
strength $\alpha/mR$ at its edge:
\bea
V(r) = \frac{\alpha}{mR}\,\delta(r-R) - \frac{\beta^2}{mR^2}\,\theta(R-r) \,,
\label{eq.new99}
\eea
with $\alpha>0$ and $\beta>0$. 
In Ref.~\cite{Habashi:2020qgw} this model was 
used to illustrate the EFT expansion for a broad resonance, 
which was reproduced in Sec.~\ref{broadresonance} using a dimer field. 
The same model had been considered in 
Ref.~\cite{Gelman:2009be} to inform the scalings of various LECs
near a narrow resonance.
We revisit the model in the context of Sec.~\ref{narrowresonance},
confirming the presence of the amplitude zero 
and providing an explicit example of convergence for low-energy observables.

\subsection{Phase shift and poles}

For the $s$ wave, the phase shift
can be obtained easily,
\begin{equation}
\cot\delta_0(k) =  - \frac{(\sqrt{k^2R^2+\beta^2} \,\cot\sqrt{k^2R^2+\beta^2} 
+ \alpha) \cot(kR) + kR}
{\sqrt{k^2R^2+\beta^2} \,\cot\sqrt{k^2R^2+\beta^2} + \alpha - kR \cot(kR)}
\, .
\label{eq.new101}
\end{equation}
We are interested in the low-energy region, $|k|\ll R^{-1}$.
If we expand this expression in powers of $kR$, the very low-energy tail
is given by the effective-range expansion. Expressions and plots for
the scattering length $a_0$ and the effective range $r_0$ can be found,
for example, in Ref.~\cite{Habashi:2020qgw}.

The model yields resonances and amplitude zeros.
A resonance appears when
$\cot\delta_0(k_\pm) -i=0$ for complex momenta $k_\pm$, which translates to
\begin{equation}
\sqrt{k_\pm^2R^2+\beta^2} \, \cot\sqrt{k_\pm^2R^2+\beta^2}=  -\alpha + ik_\pm R \,. 
\label{toyresonance}
\end{equation}
In contrast, a real zero arises from $\cot\delta_0(k_0)\to \pm \infty$, or
\begin{equation}
\sqrt{k_0^2R^2+\beta^2} \,\cot\sqrt{k_0^2R^2+\beta^2}=-\alpha +k_0R\cot(k_0R) \,. 
\label{toyzero}
\end{equation}
As pointed out in Ref. \cite{Gelman:2009be}, we expect the resonance
to be narrow when the strength of the delta-shell potential is large, 
$\alpha \gg 1$. 
In this case, one can solve Eqs. \eqref{toyresonance} and \eqref{toyzero}
as expansions in $\alpha^{-1}$. One finds a sequence of resonances
labeled by a positive integer $n=1, 2, \ldots$,
\bea
k_{Rn}R &=& n\pi
\left[1-\left(\frac{\beta}{n\pi}\right)^2-\frac{2}{\alpha} 
+\frac{3}{\alpha^2}+\frac{2}{3\alpha^3}\left(n^2\pi^2-6\right)
+{\cal O}\left(\alpha^{-4}\right) 
\right]^{1/2}\,,
\label{kR2R2toy}
\\
k_{In}R &=& \frac{(n\pi)^2}{\alpha^2} 
\left[1-\frac{3}{\alpha}+{\cal O}\left(\alpha^{-2}\right)\right] \,.
\label{kIRtoy}
\eea
Each narrow resonance is accompanied by a zero of the amplitude below it,
\begin{equation}
\left(k_{0n}R\right)^2 = \left(k_{Rn}R\right)^2 
-2 \left(\frac{n\pi}{\alpha}\right)^2
\left[1+{\cal O}\left(\alpha^{-1}\right)\right] \, .
\label{k0toy}
\end{equation}

\begin{figure}[t]
\includegraphics[scale=0.21]{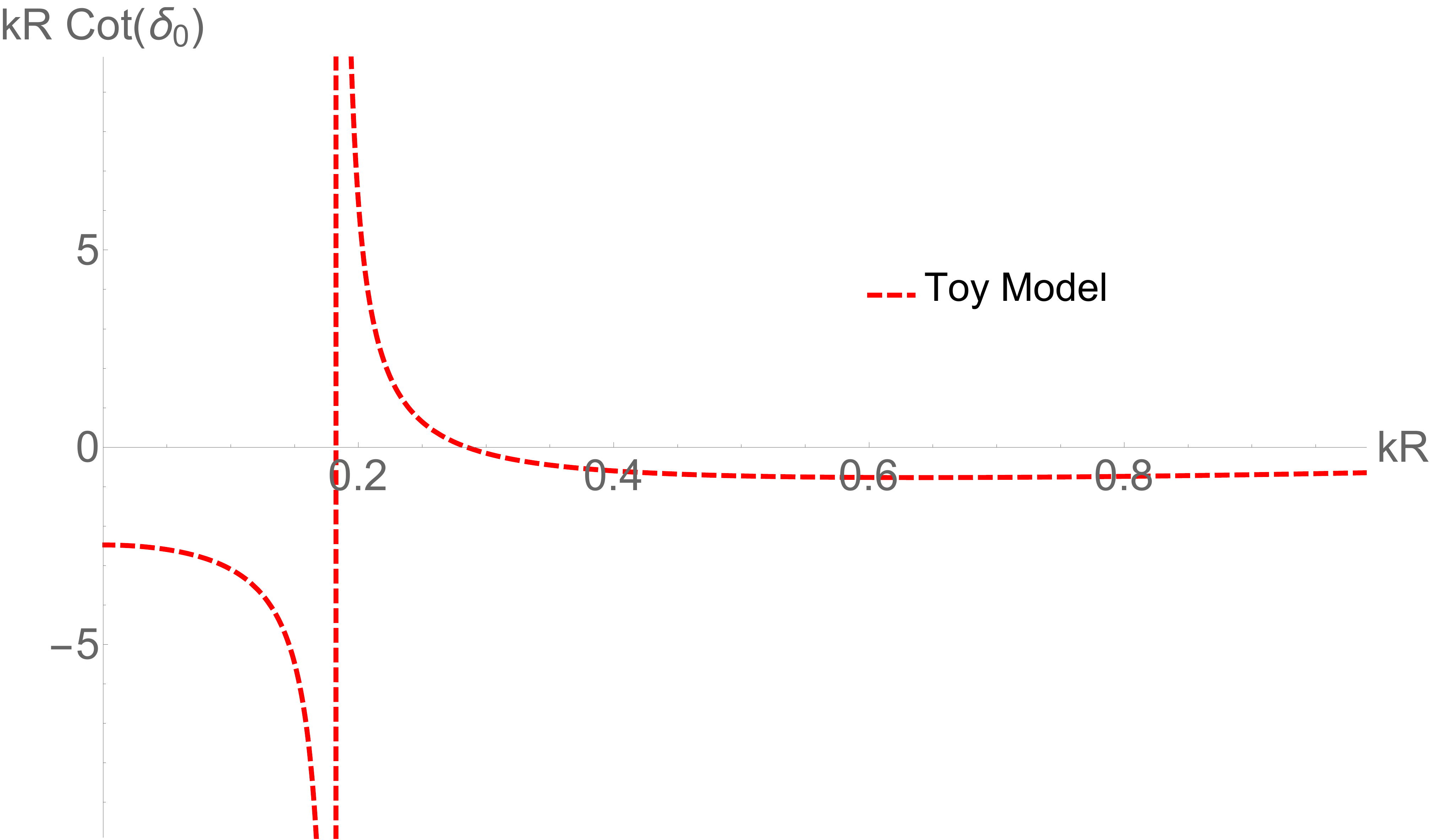}
\includegraphics[scale=0.21]{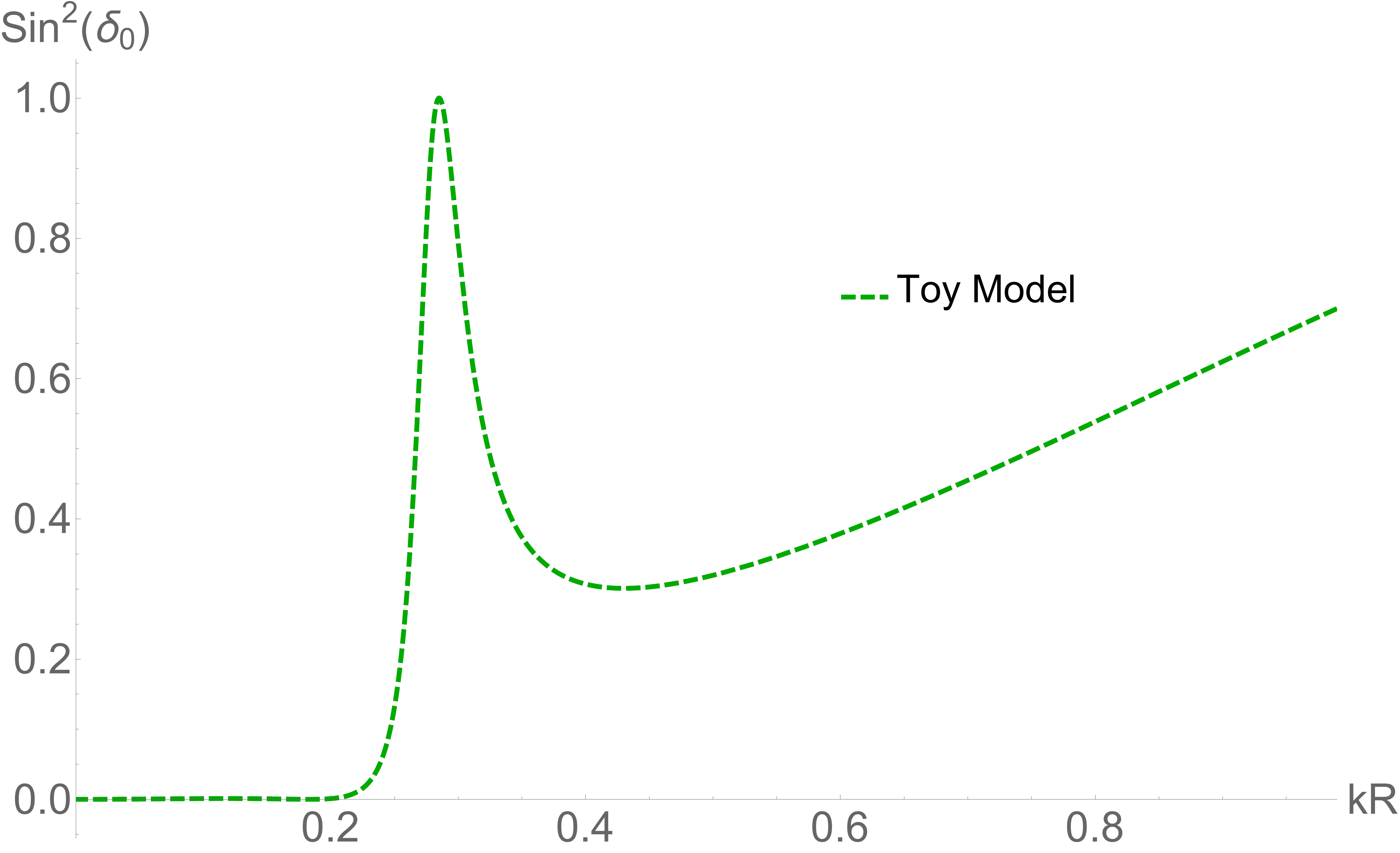}
\caption{$s$-wave phase shift for the potential \eqref{eq.new99}
with $\alpha$ and $\beta$ in Eq.~\eqref{eq.new102},
as function of the momentum in units of the inverse range ($R^{-1}$):
$kR\cot\delta_{0}$ (left panel) and $\sin^2\delta_{0}$ (right panel).
}
\label{fig.10}
\end{figure}

For specific values of $\alpha$ and $\beta$,
a resonance exists within the low-energy region. 
For the lowest narrow resonance to be in the low-energy region, 
\begin{equation}
\frac{\pi}{\alpha} \ll 1
\,, \qquad 
\pi^2-\beta^2-\frac{2\pi^2}{\alpha}
\ll 1
\,.
\label{constraint}
\end{equation}
As a concrete case, we take 
\begin{equation}
\alpha = 2 \pi^2 \,, \qquad \beta^{2} = \pi^2 - 1 \,.
\label{eq.new102}
\end{equation}
The corresponding phase shift is shown in \figref{fig.10}.
The left panel displays $kR \cot\delta_0(k)$ as a function of $kR$. 
At very small energy, the curve is approximately quadratic with
\bea
\frac{a_{0}}{R} \simeq 0.40329 \, , 
\qquad 
\frac{r_{0}}{R} \simeq -86.04272 \,. 
\label{eq.new103}
\eea
As the energy increases,
the divergence associated with the
amplitude zero becomes clearly visible at $k_0R \simeq 0.2$.
As the energy increases just a bit further, 
$\cot \delta_{0}(k)$ vanishes at $k_{r}R \simeq 0.3$.
This is the resonance region, where
a narrow resonance manifests itself as a peak in the cross section
$\sigma_{0} \propto \sin^2 \delta_{0}(k)$.
On the right panel, the peak 
around $k_{r}R \simeq 0.3$ is seen in the plot of $\sin^2 \delta_{0}(k)$ 
as a function of $kR$.
The values for the pole and zero momenta, 
$k_{\pm}$ and $k_{0}$, are listed in Table~\ref{tbl.1}. 

\begin{table}[tb]
\begin{tabular}{c||c|c|c}
& $k_{R}R$ & $k_{I}R$ & $k_{0}R$ \\
\hline
\hline
LO EFT & $0.281 \pm 0.022$ & $0.000 \pm 0.022$ & $0.18265$
\\
NLO EFT& $0.281 \pm 0.004$ & $0.022 \pm 0.004$ & $0.18265$
\\
\hline
Toy model & $0.278159$ & $0.021427$ & $0.18265$
\end{tabular}
\caption{Positions of the resonance poles ($k_{R,I}$) and 
the zero of the $T$ matrix ($k_{0}$) in units of the inverse
interaction range ($R^{-1})$.
The EFT at LO and NLO is compared with the potential \eqref{eq.new99}
with $\alpha$ and $\beta$ in Eq.~\eqref{eq.new102}. 
}
\label{tbl.1}
\end{table}

\subsection{Comparison with EFT}

We now describe this physics with the EFT of Sec.~\ref{narrowresonance}.
The toy-model parameter choice \eqref{eq.new102}
gives $k_{I}/k_{R} = 1/2\sqrt{3}\pi \simeq 1/11$
through Eqs. \eqref{kR2R2toy} and \eqref{kIRtoy}.
Since $k_{I}/k_{R} = {\cal O}(M_{lo}/M_{hi})$ in our PC,
we identify the expansion parameter as $M_{lo}/M_{hi}\sim 1/10$.
Our PC indeed captures within factors of 2 or 3
the magnitude of the various quantities
calculated in the toy model:
\begin{itemize}
\item 
$a_{0}/R = {\cal O}(1)$ and $|r_{0}|/R={\cal O}(M_{hi}^2/M_{lo}^2)$ in 
Eq. \eqref{eq.new103};
\item 
$k_0R ={\cal O}(M_{lo}/M_{hi})$ in Table~\ref{tbl.1};
\item 
$k_rR ={\cal O}(M_{lo}/M_{hi})$,
related to $a_{0}$, $r_{0}$, and $k_0$ by Eq. \eqref{eq.new51}.
\end{itemize}

\begin{figure}[t]
\includegraphics[scale=0.21]{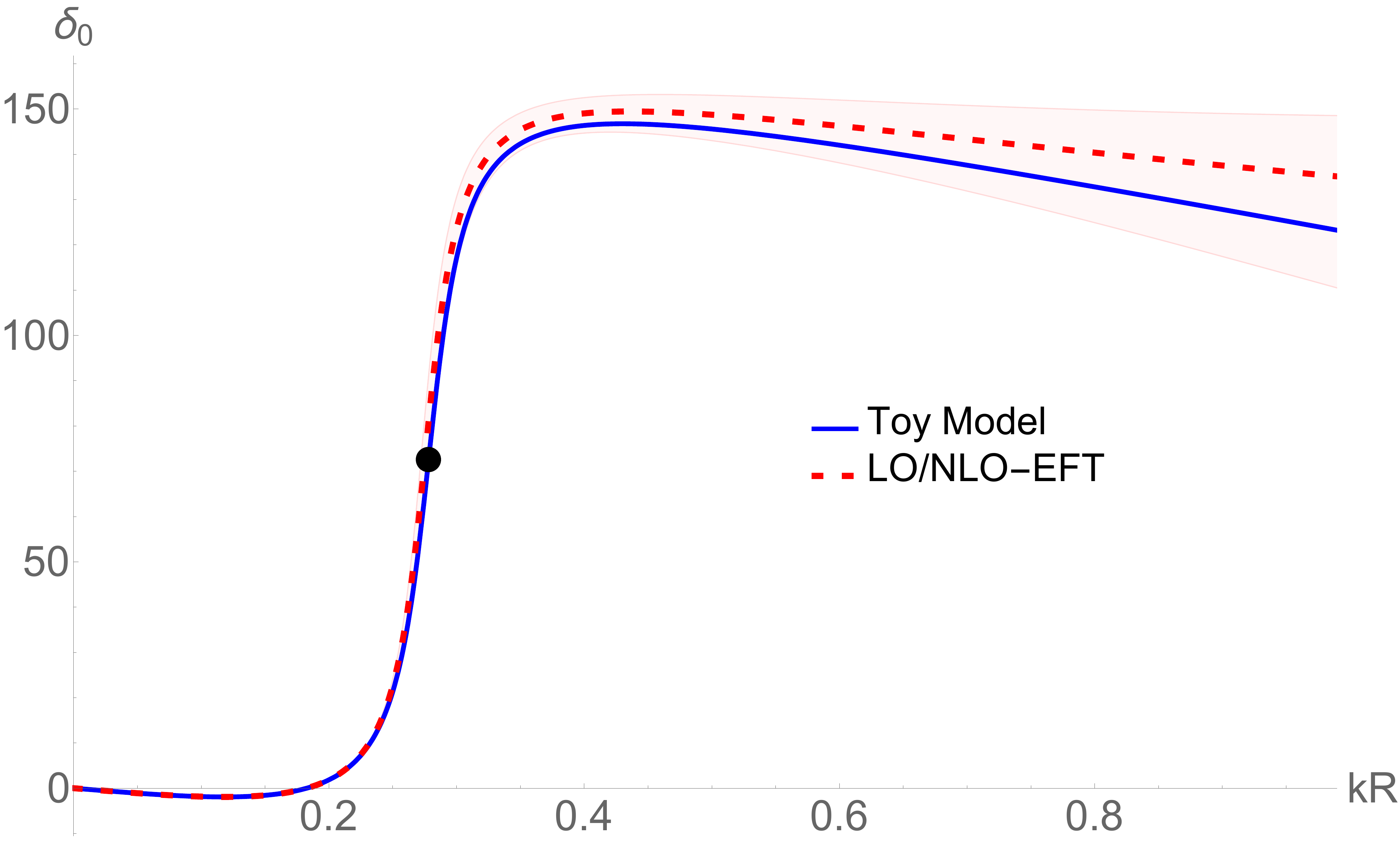}
\includegraphics[scale=0.21]{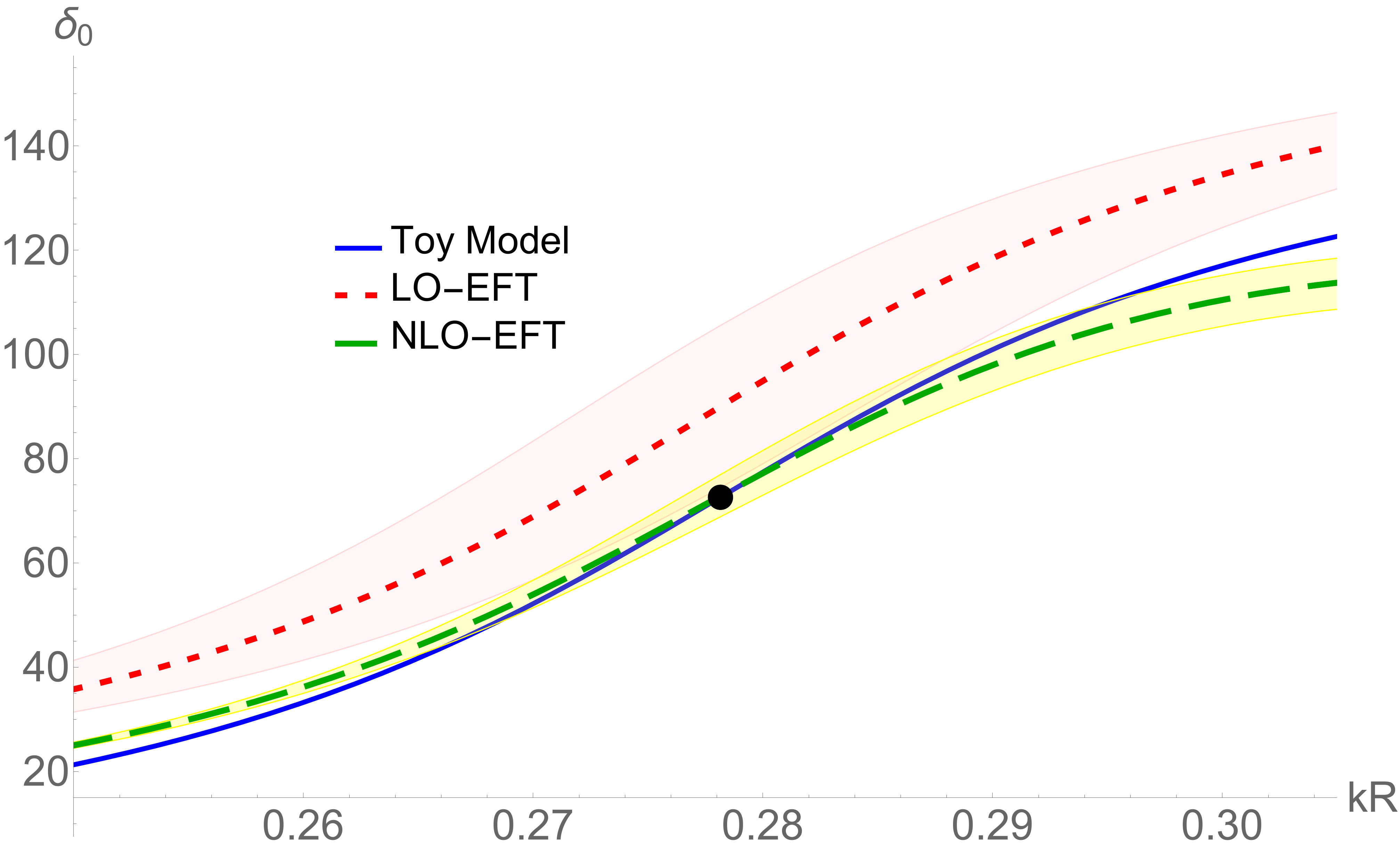}
\caption{$s$-wave phase shift $\delta_0$ in the EFT at LO and NLO 
as function of the momentum in units of the inverse range ($R^{-1}$)
compared
to the result (solid blue line) from the potential \eqref{eq.new99}
with $\alpha$ and $\beta$ in Eq.~\eqref{eq.new102}.
Left panel: 
LO/NLO phase shift (red dashed line) and error (red shadow band)
obtained from Eqs. \eqref{eq.new38} and \eqref{eq.new50}.
Right panel:
LO phase shift (red dotted line) and error (red shadow band), 
and NLO phase shift (green dashed line) and error (yellow shadow band)
from Eqs. \eqref{eq.new55} and \eqref{eq.new60}, respectively,
in a small window around the narrow resonance.
The value of the
real part of the resonance momentum is marked by a black dot
on the blue line.
}
\label{fig.11}
\end{figure}

The EFT amplitude at LO for generic low momenta, Eq.~\eqref{eq.new38}, 
has three parameters,
which we choose to fit to the position of the amplitude zero
in Table \ref{tbl.1} and
to the effective-range parameters in Eq. \eqref{eq.new103}.
At NLO the phase shift 
does not change because 
the form of $k \cot\delta_{0}(k)$ in Eq.~\eqref{eq.new50}
is the same as in LO.
The corresponding error can be estimated
from the residual cutoff dependence in $k\cot\delta_0(k)$
as $\pm \theta_{-1} k^2R^{-1}$.
The EFT phase shift, $\delta_{0}(k)$, is compared
to the toy-model phase shift on the left panel of \figref{fig.11}. 
We see that, despite all fit parameters being determined
at momenta at or below $k_0$, the NLO EFT reproduces
the toy model within error bars
throughout the low-energy region.

\begin{figure}[t]
\includegraphics[scale=0.3]{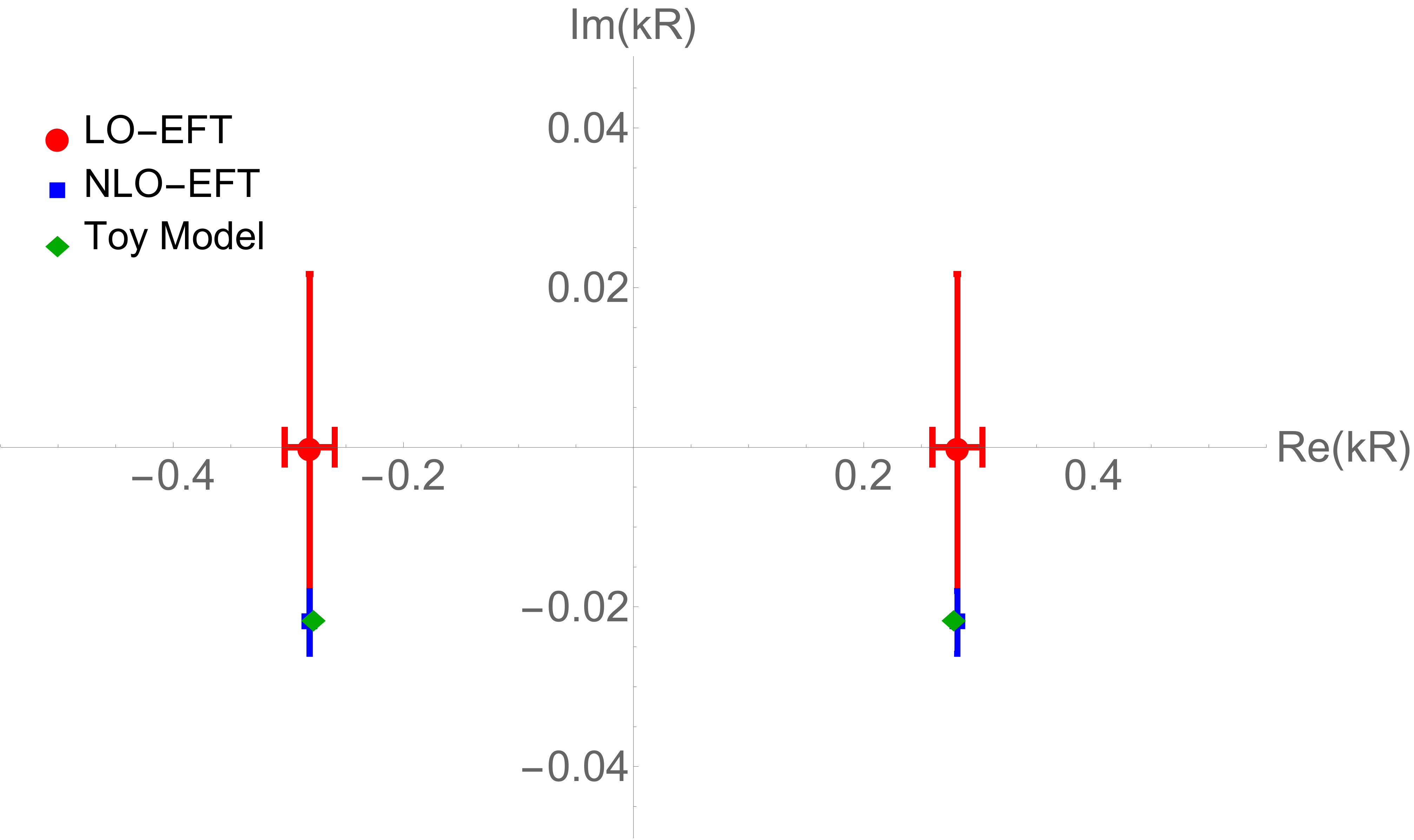}
\caption{Pole positions in the EFT at LO (red circle and error bars) 
and NLO (blue square and error bars)
compared 
to the result (green diamond)
from the potential \eqref{eq.new99}
with $\alpha$ and $\beta$ in Eq.~\eqref{eq.new102}.
}
\label{fig.12}
\end{figure}

Although the phase shift (and thus the zero position) 
is the same at LO and NLO,
the pole positions change at NLO due to the unitary term.
At LO the resonance pole is on the real axis,
Eqs.~\eqref{eq.new42} and \eqref{eq.new43}.
The LO $T$ matrix in Eq.~\eqref{eq.new38} does not contain any
residual cutoff dependence which could be used to estimate
the error in the LO pole position.
This error is ${\cal O}(M_{lo}/M_{hi})$ and could be expressed
in a number of ways in terms of the parameters appearing in
Eq.~\eqref{eq.new38}. One possible combination
is taking the average of $k_0$ and $k_r$ as representative of
$M_{lo}$, and $a_0$ as representative of $M_{hi}$, leading to an error
of magnitude
\begin{equation}
|\Delta (k_{\pm}^{(0)} R)| = \frac{a_{0}}{R} 
\left(\frac{k_{r} R + k_{0} R}{2}\right)^2 \, .
\label{eq.new105}
\end{equation}
At NLO, the resonance poles move below the real axis according
to Eqs.~\eqref{eq.new52}
and \eqref{eq.new53}.
We can translate the error from the residual cutoff dependence into
\begin{equation}
|\Delta (k_{\pm}^{(1)} R)| = 
\frac{a_{0}}{\pi R} \left(k_{r} R\right)^3 
\left[\frac{(k_{r} R)^2}{(k_{0} R)^2} - 1\right] 
\,.
\label{eq.new106}
\end{equation}
The EFT approximation for the pole positions is given in Table \ref{tbl.1}
and represented graphically in \figref{fig.12}. 
As we can see, LO and NLO results include the toy-model poles within 
their respective errors.
Both central values and errors are converging systematically,
confirming that the EFT is working properly.

Within a small window around the narrow resonance, 
the changes in pole positions are relatively large. 
An estimate of the size of the window 
is given by the magnitude of the imaginary part of the pole momentum, 
{\it i.e.} $|k - k_{R}| R \sim k_{I} R \simeq 0.02$ around $k_R$.
Resummation of the quantum corrections that produce
the resonance width leads to the LO amplitude
\eqref{eq.new55}, which contains two parameters, $k_{R}$ and $k_{I}$.
We fit them to the values from the toy model in Table~\ref{tbl.1}. 
At NLO, Eq.~\eqref{eq.new60} has an additional parameter,
$c$, which could be fitted to the phase shift at a particular
momentum within the window. 
Alternatively, we use the matching
Eqs.~\eqref{eq.new67}, \eqref{eq.new68}, and \eqref{eq.new69}
to express 
\begin{equation}
\frac{c}{R} = \frac{2k_I R}{k_{R}^2R^2 - k_{0}^2R^2 } 
\simeq  0.973707 
\label{eq.new104}
\end{equation}
from the toy-model values in Table~\ref{tbl.1}. 
As expected from our PC, $c={\cal O}(M_{hi}^{-1})$.

Although the pole positions are exact already at LO,
the phase shift changes as we go from LO to NLO.
The error at LO can be estimated from the residual cutoff dependence
just as the NLO error for generic momenta.
The residual cutoff dependence at NLO, on the other
hand, is $\propto \Lambda^{-3}$
and would underestimate the magnitude of N$^2$LO corrections.
Since nothing in principle prevents a $k^4$ term in Eq.~\eqref{eq.new60},
which would be suppressed by two powers of $M_{hi}$,
we estimate the error compared to $ck^2$ as $\pm c^2k^4/k_R$,
using $k_R$ and $c$ as proxies for $M_{lo}$ and $M_{hi}^{-1}$, respectively. 
Results are shown on the right panel of \figref{fig.11}. 
While LO works at the 20\% level,
the NLO result is very close to the toy-model phase shift.
Even though we have not fitted the latter directly,
using the matching \eqref{eq.new104} instead,
the NLO and toy-model curves intersect at $k_R$.
We see again the systematic improvement of the EFT as 
order increases.

The EFT with our PC thus describes pretty well 
the physics encoded in a narrow resonance and its concomitant
amplitude zero,
just as the PC of Sec.~\ref{broadresonance} 
does  \cite{Habashi:2020qgw} for a broad resonance with no low-energy
amplitude zero.
Next we discuss the possibility of a low-energy
amplitude zero appearing together with a broad resonance.

\section{Broad resonance with an amplitude zero}
\label{ampzero}

So far we have assumed that $C_0$ has a magnitude compatible with NDA,
with only the sizes of dimer parameters affected by the low-energy scale
$M_{lo}$. 
Different scalings of $g_0$, Eqs. \eqref{eq.new17} and \eqref{eq.new36},
lead to broad and narrow resonances, respectively.
Only for narrow resonances does the amplitude have a zero in the low-energy
region. 
In this section we consider the possibility of the existence of an
amplitude zero together with a broad resonance,
when Eqs. \eqref{eq.new10} and \eqref{eq.new17} are still expected to hold.
This requires an additional fine tuning that makes $C_0$ large,
\bea
C_{0}^{(0)}= \mathcal{O}\left(\frac{1}{M_{lo}}\right) 
\,.
\label{eq.new70}
\eea

\begin{figure}[t]
\includegraphics[trim={3.45cm 22.45cm 3.3cm 2.15cm},clip]{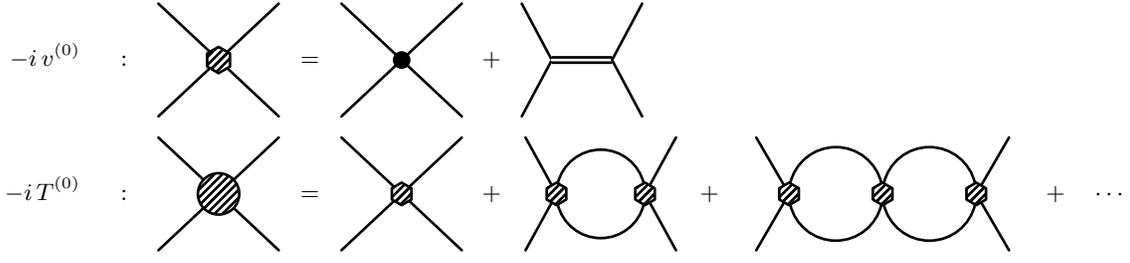}
\caption{Leading-order potential and $T$ matrix for a broad resonance 
with an amplitude zero, in terms
of the leading four-particle contact interaction (filled circle) 
and the dimer propagator (double line).}
\label{fig.7}
\end{figure}

With the assumption \eqref{eq.new70}, 
we must resum not only the dimer propagator
as in Sec. \ref{broadresonance},
but also the non-derivative contact interaction.
The corresponding diagrams in \figref{fig.7} lead to
\bea
T^{(0)}(k) & = & \frac{4\pi}{m}\,
\frac{C_{0}^{(0)} + g_{0}^{(0)2} D_0^{(0)}(k)}
{1 + (C_{0}^{(0)} + g_{0}^{(0)2} D_0^{(0)}(k)) I_{0}(k)}  
\label{eq.new71}\\
& = & 
-\frac{4 \pi}{m} \left(-\frac{1}{a_0} +\frac{r_0k^2}{2} \frac{1}{1-k^2/k_{0}^2} 
- ik - L_{-1}k^2 +\ldots \right)^{-1} 
\,,
\label{eq.new72}
\eea
where the three LO LECs renormalize the amplitude with 
\bea
\Delta^{(0)}(\Lambda) & = & \frac{k_{0}^2}{m} \,
\frac{L_{1} -1/a_0}{L_1-1/a_0-r_0k_0^2/2} \,,
\label{eq.new73}\\
g_{0}^{(0)}(\Lambda)  & = & 
\left[-\frac{r_0k_{0}^4}{2m(L_1-1/a_0-r_0k_0^2/2)^2}\right]^{1/2}\,,
\label{eq.new74}\\
C_{0}^{(0)}(\Lambda)  & = &  -\frac{1}{L_1-1/a_0-r_0k_0^2/2}\,.
\label{eq.new75}
\eea
The amplitude is of the same form as the NLO amplitude \eqref{eq.new50} 
for a narrow resonance with the replacement $k_r^2\to k_0^2(1+a_0r_0k_0^2/2)^{-1}$.
However, the amplitude \eqref{eq.new72} holds throughout the low-energy region,
$a_0$ and $r_0$
being identified as before with the scattering length and effective range. 
Again, renormalization requires $r_0<0$ since
$k_{0}^4 > 0$ must be real for $g_{0}^{(0)}(\Lambda)$ also to be real.
The amplitude-zero location $k_{0}$ can be real or imaginary, 
but $|k_0|={\cal O}(M_{lo})$.
As in the amplitude \eqref{eq.new20},
$|a_0|\sim |r_0| ={\cal O}(1/M_{lo})$.
The difference lies in the other effective-range
parameters, which are determined by $M_{lo}$
instead of $M_{hi}$. For example, the shape parameter at LO is
\cite{SanchezSanchez:2017tws}
\begin{equation}
P_0^{(0)}=-\frac{4}{r_0^2k_0^2}
\,.
\label{eq.new76}
\end{equation}
As expected, in the limit $|k_0|\to \infty$ Eq. \eqref{eq.new72}
reduces to Eq. \eqref{eq.new20}. At the same time, $C_0^{(0)}\to 0$, and 
Eqs. \eqref{eq.new73} and \eqref{eq.new74} go into 
Eqs. \eqref{eq.new21} and \eqref{eq.new22}, respectively.

The amplitude \eqref{eq.new71}
was considered with dimensional regularization and
minimal subtraction in Ref. \cite{Kaplan:1996nv},
while renormalization equivalent to that above was performed
in Ref. \cite{Braaten:2007nq} starting from a sharp momentum cutoff.
This amplitude is frequently invoked in discussions
of Feshbach resonances --- see for example Ref. \cite{Diehl:2005an}.
In this context $C_0$ is usually considered a natural, small background,
for which resummation is actually unnecessary. 
The possibility of a fine-tuned $C_0$, resulting in a low-energy zero,
was investigated in Ref. \cite{SanchezSanchez:2017tws} with an eye on the 
two-nucleon $^1S_0$ channel.
In Ref. \cite{Rupak:2018gnc} the interest was 
the motion of poles in the three-nucleon system:
neutron-deuteron scattering in the spin-$3/2$ channel shows a shallow 
amplitude zero
below threshold, which moves into the scattering region as 
the two-nucleon system approaches unitarity. 
The focus of all these references was on shallow bound and/or virtual 
states.

Here we are interested in Eq. \eqref{eq.new72} for the description
of a resonance in the presence of the zeros at $k=\pm k_0$.
In order to find the poles we rearrange the 
denominator of the $T$ matrix in Eq.~\eqref{eq.new72} as
\begin{equation}
 k^3 -i\, \frac{k^2}{a_{0}} \left(1+\frac{a_0r_0k_{0}^2}{2}\right) - k k_{0}^2
+i\, \frac{k_{0}^2}{a_0}  =0 \,.
\label{eq.new77}
\end{equation}
For real values of $a_{0}$, $r_{0}$, and $k_{0}^2$ this 
cubic polynomial has a symmetry relative to the imaginary 
axis: if $k$ is a root, so is $- k^\ast$. 
Hence the roots of Eq.~\eqref{eq.new77} should either 
be imaginary or come in pairs symmetric with respect to the 
imaginary axis. 
Since we have only three 
roots there are only two possibilities: 
{\it i)} all the three roots are imaginary \cite{SanchezSanchez:2017tws};
{\it ii)} one root is imaginary and the other two 
are symmetric relative to the imaginary axis, 
\bea
k_{\pm}^{(0)} & = & \pm k_{R}^{(0)} - ik_{I}^{(0)} \, , 
\label{eq.new7879}\\
k_{3}^{(0)} & = & i \kappa^{(0)} \, , 
\label{eq.new80}
\eea
with real $k_{R,I}^{(0)}$ and $\kappa^{(0)}$ satisfying the conditions
\bea
\kappa^{(0)} - 2 k_{I}^{(0)} & = & 
\frac{1}{a_{0}} + \frac{r_{0}}{2} k_{0}^2 \, , 
\label{eq.new82}\\
k_{R}^{(0)2} + k_{I}^{(0)2} - 2 \kappa^{(0)} k_{I}^{(0)} & = & k_{0}^2 \, , 
\label{eq.new83}\\
\kappa^{(0)} \left(k_{R}^{(0)2} + k_{I}^{(0)2}\right) 
& = & \frac{k_{0}^2}{a_{0}} \, .
\label{eq.new84}
\eea
From these relations,
\begin{equation}
k_{I}^{(0)} \left[ k_{R}^{(0)2} 
+ \left(\kappa^{(0)} - k_{I}^{(0)} \right)^2\right]
=- \frac{r_{0}\,k_{0}^4}{4} 
\, ,
\label{eq.new85}
\end{equation}
which shows that $k_{I}^{(0)}>0$ for $r_0<0$.
As before, the resonance 
poles are in the lower half of the complex momentum plane
thanks to the renormalization constraint on the effective range
for a non-ghost dimer field.
In addition to the two resonance poles,
the amplitude zeros induce an extra pole on the imaginary axis.
In the $|k_{0}|\to\infty$ limit,
\begin{equation}
\lim_{|k_{0}|\to\infty}\,\kappa^{(0)} = \frac{r_{0}}{2} k_{0}^2 \, , 
\label{eq.new96}
\end{equation}
while $k_{I,R}$ are given by Eqs.~\eqref{eq.new23} and \eqref{eq.new24}.
Thus, for $|k_0|={\cal O}(M_{hi})$ the additional pole represents a deep
virtual (bound) state for $k_0$ real (imaginary).
For the case considered here, $|k_0|={\cal O}(M_{lo})$,
it lies instead within the region of applicability of the theory.
The $S$ matrix now takes the form \eqref{eq.new86} with $N=3$ ($n=\pm, 3$) 
and 
\begin{equation}
\phi^{(0)}(k) = 0 \,. 
\label{eq.new87}
\end{equation}

We can proceed to NLO as in the previous cases.
The residual cutoff dependence of the LO amplitude is the same as in 
Eq.~\eqref{eq.new26}, indicating that LECs exist at NLO to remove it.
A quick calculation shows that $g_{2}$ does not accomplish this task
in perturbation theory,
and we can surmise that it is suppressed by the natural
two powers of $M_{hi}$ compared to $g_0$.
As in the case of a single shallow bound or virtual state 
\cite{vanKolck:1997ut,Kaplan:1998tg,Kaplan:1998we,vanKolck:1998bw},
an enhanced $C_0$ requires an enhanced $C_2$ for renormalization at NLO.
In addition to the LECs \eqref{eq.new27}, we need
\bea
C_{0}^{(1)} = \mathcal{O}\left(\frac{1}{M_{hi}}\right) 
\,,\qquad 
C_{2}^{(1)} = \mathcal{O}\left(\frac{1}{M_{lo}^2M_{hi}}\right)
\,.
\label{eq.new88}
\eea
The new interaction with LEC $C_{2}$ means that an additional parameter
is introduced at NLO, which we can associate with a correction to
the shape parameter \eqref{eq.new76}.

\begin{figure}[t]
\includegraphics[trim={2.9cm 24.25cm 2.9cm 2.15cm},clip]{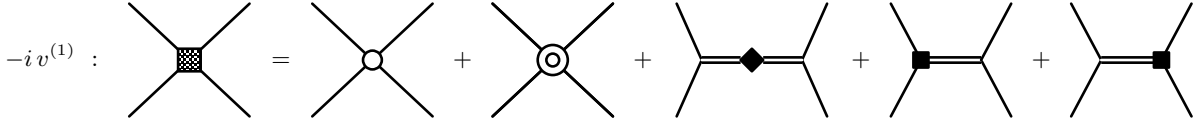}
\caption{Next-to-leading-order potential from
the four-particle no-derivative (circle)
and two-derivative contact interactions (circled circle),
dimer residual mass (solid diamond), 
and two-particle/dimer coupling (solid square).
Other symbols as in \figref{fig.1}.}
\label{fig.8}
\end{figure}

\begin{figure}[t]
\includegraphics[trim={3.2cm 24.25cm 3.5cm 2.15cm},clip]{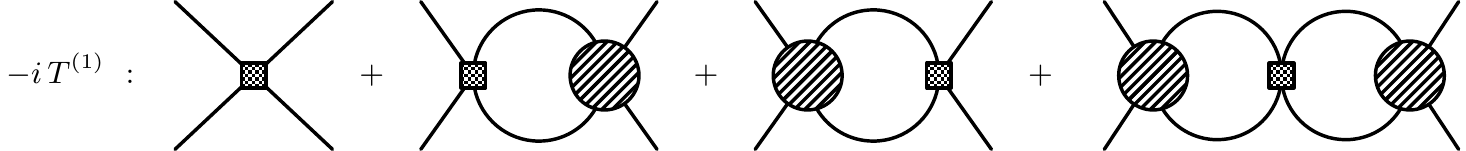}
\caption{Next-to-leading-order $T$ matrix for a broad resonance
in terms of the NLO potential (shaded square, \figref{fig.8})
and the LO $T$ matrix (\figref{fig.7}).}
\label{fig.9}
\end{figure}

These corrections lead to the NLO potential and amplitude shown in 
Figs. \ref{fig.8} and \ref{fig.9},
which are the analogs of Figs. \ref{fig.2} and \ref{fig.3}, respectively.
They lead to
\bea
\frac{4\pi}{m} 
\left(\frac{T^{(1)}}{T^{(0)}{}^2} - \delta T^{(1)-1} \right) 
&=& 
\left[C_{0}^{(1)}\left(L_{1}-\frac{1}{a_0}\right)+ C_{2}^{(1)}L_{3}
- g_{0}^{(1)}\sqrt{-2mr_0}\right]\left(L_{1}-\frac{1}{a_0}\right)
-\frac{mr_0}{2}\Delta^{(1)}
\notag\\
&& 
+\bigg\{C_{0}^{(1)} \left(L_{1}-\frac{1}{a_0}\right)
+\frac{C_{2}^{(1)}}{2}
\left[L_{3}+\frac{2}{r_0}\left(L_{1}-\frac{1}{a_0}\right)^2\right]
\notag\\
&& 
-\frac{g_{0}^{(1)}}{k_{0}^2}\sqrt{-\frac{2m}{r_0}}
\left(L_{1}-\frac{1}{a_0}+\frac{r_0k_{0}^2}{2}\right)
- \frac{m}{k_{0}^2}\Delta^{(1)}
- \frac{L_{-1}}{r_0}\bigg\}
\frac{r_0k^2}{1-k^{2}/k_{0}^2}
\notag\\
&&+ \biggl[ \frac{C_{0}^{(1)}}{4}
-\frac{C_{2}^{(1)}}{r_0^2k_0^2}
\left(L_1-\frac{1}{a_0}\right)\left(L_1-\frac{1}{a_0}-r_0k_{0}^2\right)
\notag\\
&& 
- \frac{g_{0}^{(1)}}{2 k_0^2} \sqrt{-\frac{2m}{r_0}}
- \frac{m}{2 r_0k_0^4}\Delta^{(1)}
+ \frac{L_{-1}}{r_0^2k_0^2}
\biggr]
\frac{r_0^2k^4}{1-k^{2}/k_{0}^2}
\notag \\
&&+ \biggl[
\frac{C_0^{(1)}}{4 k_{0}^2}+ \frac{C_{2}^{(1)}}{4}
- \frac{g_{0}^{(1)} }{2 k_0^4}\sqrt{-\frac{2m}{r_0}}
- \frac{m}{2 r_0 k_{0}^6}\Delta^{(1)}
\biggr] \frac{r_0^2k^6}{(1-k^{2}/k_{0}^2)^2} 
\notag \\
&& - L_{-3}\,k^{4} + \ldots 
\notag \\
&=& -\left(P_0+\frac{4}{r_0^2k_0^2}\right) 
\left(\frac{r_0}{2}\right)^3\frac{k^{4}}{1-k^{2}/k_{0}^2} + \ldots \,,
\label{eq.new89}
\eea
where we imposed the renormalization conditions that no changes
are induced in the energy dependence and parameters of the LO amplitude,
other than the presence of a new term proportional to $P_0-P_0^{(0)}$.
This is accomplished with
\bea
C_{2}^{(1)}(\Lambda) & = &  \frac{k_0^2}{(L_1-1/a_0 -r_0k_{0}^2)^2} 
\left(\frac{r_0}{2}\right)^3
\left(P_0+\frac{4}{r_0^2k_0^2}+\frac{8L_{-1}}{r_0^3k_0^2}\right)
\,,
\label{eq.new90} \\
C_{0}^{(1)}(\Lambda) & = & -\frac{k_0^2}{(L_1-1/a_0 -r_0k_{0}^2)^3} 
\left(\frac{r_0}{2}\right)^3
\bigg\{\left(P_0 +\frac{4}{r_0^2k_0^2}\right)
\left[L_3-k_0^2L_1+\frac{k_0^2}{a_0}\left(1-\frac{a_0r_0k_0^2}{2}\right)\right]
+\frac{8L_{-1}}{r_0^3k_0^2}
\left(L_3-r_0k_0^4\right)
\bigg\}\,,
\notag\\
\label{eq.new91} \\
g_{0}^{(1)}(\Lambda) & = & 
-\left(-\frac{1}{2mr_0}\right)^{1/2}\frac{1}{(L_1-1/a_0 -r_0k_{0}^2)^3} 
\left(\frac{r_0k_0}{2}\right)^4
\,
\bigg\{\left(P_0+\frac{4}{r_0^2k_0^2}\right)
\left[L_3-\frac{2}{r_0}\left(L_1-\frac{1}{a_0}\right)
\left(L_1-\frac{1}{a_0}+r_0k_0^2\right)\right]
\notag\\
&&
\qquad\qquad\qquad\qquad\qquad\qquad\qquad\qquad\qquad\qquad\quad
+\frac{8L_{-1}}{r_0^3k_0^2}
\left[L_3-2k_0^2L_1+\frac{2 k_0^2}{a_0}\left(1-\frac{a_0r_0k_0^2}{4}\right)\right]
\bigg\}
\,,
\label{eq.new92}\\
\Delta^{(1)}(\Lambda) & = & -
\frac{L_1-1/a_0}{m(L_1-1/a_0 -r_0k_{0}^2)^3} 
\left(\frac{r_0k_0^2}{2}\right)^3
\left[\left(P_0+\frac{4}{r_0^2k_0^2}\right)\left(L_1-\frac{1}{a_0}\right)
+\frac{4L_{-1}}{r_0^2} \right]
\,.
\label{eq.new93} 
\eea

The NLO amplitude is then 
\begin{equation}
T^{(0+1)}(k)= -\frac{4 \pi}{m}\left[-\frac{1}{a_{0}} + \frac{r_{0}}{2}\,k^2
-\left(\frac{r_0}{2}\right)^3 \frac{P_0 k^4}{1-k^2/k_{0}^2} - ik \right]^{-1} 
+ \ldots \,.
\label{eq.new94}
\end{equation}
This result was obtained in Ref. \cite{SanchezSanchez:2017tws}
using two dimer fields, but we see here that no such
complication is necessary.

The NLO corrections induce changes in the pole positions,
\bea
k_{n}^{(0+1)} = k_{n}^{(0)} 
\left[1
+ i \left(P_0+\frac{4}{r_0^2k_0^2}\right) 
\left(\frac{r_{0}k_{n}^{(0)}}{2}\right)^3
\frac{1-k_{n}^{(0)2}/k_{0}^2}{(1-k_{n}^{(0)2}/k_{0}^2)^2 + i r_{0}k_{n}^{(0)}} 
\right]\,,
\label{eq.new95}
\eea
and the background phase,
\bea
\phi^{(0+1)}(k) & = & - 
\left(P_{0} + \frac{4}{r_{0}^2\,k_{0}^2}\right) 
\left(\frac{r_{0}}{2}\right)^3 k_{0}^2 k \,.
\label{eq.new97}
\eea
In the limit $|k_{0}|\to\infty$ we regain 
Eqs. \eqref{eq.new33} and \eqref{eq.new34} for the two poles $n=\pm$.
However, for $|k_{0}|={\cal O}(M_{hi})$ the perturbative expansion for 
the $n=3$ pole does not converge. As a consequence, the expansion
for the phase does not commute with the $|k_{0}|\to\infty$ limit, and 
the three-pole $\phi^{(0+1)}(k)$ in Eq. \eqref{eq.new97}
does not go into the two-pole Eq. \eqref{eq.new35}.

Like in the case without a zero, the ``\ldots'' in Eq. \eqref{eq.new94}
contain terms $\propto \Lambda^{-3}$ indicating that
new interactions appear at or before N$^3$LO,
{\it e.g.} $g_2$ at N$^2$LO.
We again stop at NLO although the procedure can be continued 
to higher orders if there is interest.
There is no apparent obstacle to a consistent EFT formulation 
of a broad resonance with a low-energy amplitude zero based
on the PC presented here. 
Yet, we have been unable to find an explicit potential-model realization
of this idea. For example, we found no combination of 
$\alpha$ and $\beta$ in the potential \eqref{eq.new99} that
yields both a broad resonance and an amplitude zero.
This is likely because there are not enough parameters to tune,
but it leaves open the question of if (and how) this scenario
can be realized.

\section{Conclusion}
\label{conclusion}

The existence of resonant poles in the scattering of two nonrelativistic
particles requires a nonperturbative treatment of the dominant interactions
between these particles, while a systematic expansion 
of the amplitude relies on distorted-wave perturbation theory being applied
to subleading interactions.
Following Refs. \cite{Weinberg:1962hj,Kaplan:1996nv},
we investigated here a formulation of the low-energy effective
field theory where 
the resonance poles enter through the propagation of a particle (dimer)
with the resonance quantum numbers, described by its own field
in the theory's Lagrangian. 
The leading-order amplitude is indeed nonperturbative in a vicinity
of the pole. In the case of a narrow resonance this region is small, and
outside perturbation theory holds.
A narrow resonance is naturally accompanied by a low-energy zero of
the amplitude, which falls outside the nonperturbative region.
In contrast, for a broad resonance the leading-order amplitude is
nonperturbative in most of the region where the EFT applies.
A low-energy zero of the amplitude appears only
upon fine tuning.

We have explicitly constructed manifestly renormalized amplitudes for 
both broad and narrow resonances up to next-leading order in the EFT expansion.
The case of a broad resonance without amplitude zero 
gives the amplitude obtained with only momentum-dependent interactions
in Ref. \cite{Habashi:2020qgw}.
When the zero is present, we provided an alternative
derivation of the amplitude from Ref. \cite{SanchezSanchez:2017tws},
which employed two dimer fields and focused on three poles
with imaginary momenta instead of a resonance.
For a narrow resonance, we corrected the derivation 
of Ref. \cite{Alhakami:2017ntb}, which itself 
improved on Refs. \cite{Bedaque:2003wa,Gelman:2009be}.
It would be interesting to build the equivalent EFT without a dimer
field in the two cases where a zero is present.

The sign of the kinetic term for the dimer field
is linked to the sign of the effective range.
It has been shown 
\cite{vanKolck:1997ut,Kaplan:1998tg,Kaplan:1998we,vanKolck:1998bw}
that positive effective range can be handled in an EFT 
with a ghost auxiliary field, 
as long as its kinetic term is treated perturbatively.
The amplitude then has a single pole and cannot accommodate resonances.
In all resonant cases, we took the kinetic term 
to have the usual sign associated with a positive-norm state.
It is remarkable that, as for purely 
momentum-dependent interactions \cite{Habashi:2020qgw},
renormalization of the leading-order amplitude requires 
the resonance poles to be on the lower half
of the complex momentum plane.
This is what is expected from the Wigner bound
\cite{Wigner:1955zz} on the rate of 
change of the phase shift with respect to the energy for a finite-range,
energy-independent potential:
the range of the potential
imposes an upper bound on the 
effective range \cite{Fewster:1994sd,Phillips:1996ae},
which restricts resonance positions (if they exist). 

The subleading amplitudes can then be renormalized in 
(distorted-wave) perturbation theory and lead to small
changes in the pole positions.
We confirmed the improvement at next-to-leading order in the case
of a narrow resonance using an explicit toy potential 
posing as underlying theory \cite{Gelman:2009be}. 
The same had previously been done for a broad resonance without amplitude zero
\cite{Habashi:2020qgw}, and we did not find a combination
of potential parameters that gives rise to a low-energy amplitude zero.

We hope these ideas find application in the enormously rich physics
of nuclear reactions, where resonances abound. 
In most situations, the Coulomb interaction must be added.
Work in this direction is in progress, so as to extend the results
of Refs. \cite{Higa:2008dn,Gelman:2009be} to more general cases.
(For related work on the interplay between finite (effective) range
and Coulomb interactions, see 
Refs. \cite{Barford:2002je,Schmickler:2019ewl,Luna:2019ufu}.)

\section*{Acknowledgments}

UvK thanks Hans-Werner Hammer for discussions on
the resummation of a negative dimer kinetic term. 
This research was supported in part
by the U.S. Department of Energy, Office of Science,
Office of Nuclear Physics, under award number DE-FG02-04ER41338.

\end{document}